\documentclass[sigconf]{acmart}
\makeatletter
\let\footnotetextcopyrightpermission\@gobble 
\makeatother
\AtBeginDocument{%
  }

\theoremstyle{acmplain}
\newtheorem{theorem}{Theorem}
\newtheorem{lemma}{Lemma}
\newtheorem{corollary}{Corollary}
\newtheorem{proposition}{Proposition}
\newtheorem{definition}{Definition}
\newtheorem{remark}{Remark}
\newtheorem{example}{Example}
\newtheorem{assumption}{Assumption}

\newcommand{\runin}[1]{\noindentparagraph{\normalfont\bfseries #1}}

\makeatletter
\newcommand{\runinfixed}[1]{\par\addvspace{.5\baselineskip}%
  \noindent{\normalfont\bfseries #1}}
\makeatother

\DeclareMathOperator*{\argmax}{arg\,max}
\newcommand{\Wel}{\mathrm{Wel}}
\newcommand{\Rev}{\mathrm{Rev}}
\newcommand{\Zbc}{\mathcal{Z}^{\mathrm{BC}}}
\newcommand{\POW}[1]{\big[#1\big]^{+}}

\usepackage{colortbl}
\usepackage{placeins}
\definecolor{BestC}{HTML}{D6EAF8}
\definecolor{SecondC}{HTML}{FDEBD0}
\definecolor{headgray}{HTML}{ECEFF1}
\definecolor{gaincol}{HTML}{1E8449}
\definecolor{costcol}{HTML}{C0392B}
\newcommand{\bfirst}[2]{\cellcolor{BestC}\textbf{#1}#2}
\newcommand{\usecond}[1]{\cellcolor{SecondC}\underline{#1}}
\newcommand{\gain}[1]{\textcolor{gaincol}{\textbf{#1}}}
\newcommand{\cost}[1]{\textcolor{costcol}{\textbf{#1}}}

\setcopyright{acmlicensed}
\copyrightyear{2027}
\acmYear{2027}

\acmConference[WWW '27]{The ACM Web Conference 2027}{Submission}{\#2029}

\begin{document}

\title{Certified Boost Design for Quality-Aware Auto-Bidding}

\author{Huanyu Yan}
\authornote{Both authors contributed equally to this research.}
\email{huanyuyan@link.cuhk.edu.cn}
\orcid{0000-0001-9619-066X}
\author{Yu Huo}
\authornotemark[1]
\email{yuhuo@link.cuhk.edu.cn}
\orcid{0009-0007-6642-0314}
\affiliation{%
  \institution{The Chinese University of HongKong, Shenzhen}
  \city{Shenzhen}
  \state{Guangdong}
  \country{China}
}

\author{Min Lu}
\affiliation{%
  \institution{Tencent Technology (Shenzhen) Co., Ltd.}
  \city{Shenzhen}
  \country{China}}
\email{alfredolv@tencent.com}
\orcid{0009-0005-6702-4263}

\author{Weitong Ou}
\affiliation{%
 \institution{Tencent Technology (Shenzhen) Co., Ltd.}
  \city{Shenzhen}
  \country{China}}
\email{wendyou@tencent.com}
\orcid{0000-0001-9088-621X}

\author{Xingyan Shi}
\affiliation{%
 \institution{The Chinese University of HongKong, Shenzhen}
 \city{Shenzhen}
 \state{Guangdong}
 \country{China}}
\email{xingyanshi@link.cuhk.edu.cn}
\orcid{0009-0003-4901-3013}

\author{Ruihe Shi}
\affiliation{%
  \institution{The Chinese University of HongKong, Shenzhen}
  \city{Shenzhen}
  \country{China}}
\email{ruiheshi@link.cuhk.edu.cn}
\orcid{0009-0003-1522-7865}

\author{Xiaoying Tang}
\correspondingauthor
\affiliation{%
  \institution{The Chinese University of HongKong, Shenzhen}
  \city{Shenzhen}
  \country{China}}
\email{tangxiaoying@cuhk.edu.cn}
\orcid{0000-0003-3955-1195}

\renewcommand{\shortauthors}{Yan et al.}

\begin{abstract}
  Online advertising platforms balance advertiser value, publisher-assessed ad quality, and platform revenue. However, boost design in this three-stakeholder architecture remains insufficiently studied. In this paper, we develop certified boost mechanisms under a payment rule that deducts boosts from winners' payments. We establish a capped benchmark-competitive condition that guarantees a quality-aware liquid-welfare lower bound parameterized by the certificate strength, boost cap, and bid cap, with an explicit penalty for an insufficient boost cap. We then derive a closed-form canonical boost. For the observed state, it matches the welfare-optimal benchmark allocation whenever its certificate margin seats every benchmark winner, and among all certified boosts that do so it maximizes the platform's retained revenue. On AuctionNet, a public benchmark built on Alibaba's real-world production advertising environment, we evaluate canonical across ten bidding policies on multiple episodes. Under a twofold shock to the strongest rival's bid, canonical achieves $97.6\%$ mean welfare efficiency, compared with $91.1\%$ for the strongest baselines. Controlled stress tests link this protection to the certificate margin, while a ten-policy cap sweep shows diminishing welfare gains as retained revenue falls.
\end{abstract}


\ccsdesc[500]{Information systems~Computational advertising}
\ccsdesc[300]{Information systems~Online advertising}
\ccsdesc[300]{Theory of computation~Algorithmic mechanism design}
\ccsdesc[100]{Computing methodologies~Reinforcement learning}

\keywords{online advertising, auto-bidding, boosted auctions, publisher
  quality}


 \maketitle

\section{Introduction}
Online advertising is allocated through real-time auctions at web scale, and advertisers increasingly bid through automated agents that optimize under budget and return-on-investment constraints \cite{zhang2025adapting,ou2023optimal}. A critical challenge for advertising platforms lies in designing auction rules that balance efficiency and revenue. One prominent approach involves platform-determined boost factors, which may take additive, multiplicative, or hybrid forms. Specifically, an additive boost factor incorporates an additional term during the auction process, which is combined with the advertiser’s commercial value to determine the winning bidder. Existing research on boost factors demonstrates that a well-designed boost mechanism can significantly improve auction efficiency and social welfare \cite{hummel2016machine,golrezaei2017boosted,golrezaei2021boosted,deng2021towards,balseiro2021robust,likhodedov2004methods,sandholm2015automated}.

\begin{figure}
    \centering
    \includegraphics[width=\linewidth]{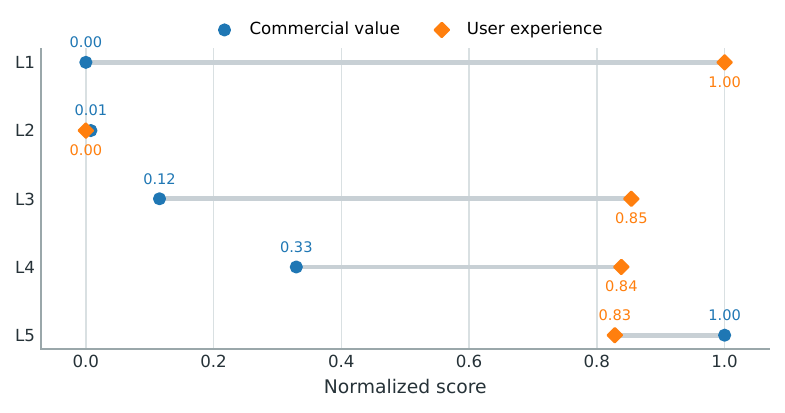}
    \caption{Normalized commercial value and user experience by ARPU segment (L1: Lowest to L5: Highest). Each line connects the two scores for one segment.}
    \Description{Paired horizontal dot plot for ARPU segments L1 through L5, with blue commercial-value dots and orange user-experience dots connected by gray lines.}
    \label{fig:value_quality_diff}
\end{figure}

However, incorporating publisher-assessed quality value introduces new theoretical challenges.  This dimension represents a publisher's assessment of an advertisement's suitability for their platform, considering factors such as brand alignment, content appropriateness, and audience perception. It represents the trade-off between potential advertising revenue and user experience, where publishers may reject ads deemed incompatible with their brand image (e.g., premium platforms excluding low-cost product ads). Figure~\ref{fig:value_quality_diff} presents empirical data from a large commercial advertising platform, illustrating the relationship between normalized commercial conversion profits and user experience scores across different ARPU (Average Revenue Per User) segments. Notably, the analysis reveals no positive correlation between these two dimensions, highlighting the need for incorporating additional quality metrics to adequately address publisher concerns regarding user experience. Besides, major online advertising systems (e.g., Google Ads \cite{google_ads_quality}) systematically incorporate ads quality scores to balance monetization with user experience and platform users' experience.

After incorporating quality considerations, the design of boost factors faces the fundamental challenge of misaligned interests among the three stakeholders. Specifically, advertisers primarily focus on commercial metrics such as conversion rates when placing ads \cite{duan2025adaptable,xu2024auto,ma2025genauction}, while largely disregarding the potential impact of ad displays on publisher traffic, whereas publishers value both commercial and quality dimensions \cite{google_ads_quality,yuan2014analyzing,yoon2010optimal,awaya2025quality,camboni2025bidding,huang2019empirical,wang2020quality}. This divergence creates conflicting evaluation standards, necessitating auction mechanisms that can properly balance these competing objectives. Consequently, traditional bidding efficiency analyzes based solely on commercial value and assumed stakeholder alignment become inapplicable.

In this work, we address this three-stakeholder challenge by investigating boost designs that reconcile commercial value, publisher quality, and platform revenue. We adopt the liquid welfare setting~\cite{dobzinski2014efficiency,lu2017liquid} and show that, conditional on the certificate margin seating the welfare-optimal winner, the capped benchmark-competitive set admits a canonical boost that maximizes retained revenue among all such certified boosts. The platform's design problem therefore reduces to choosing three scalar parameters: the certificate strength, the boost-cap level, and the bid cap. The platform observes the submitted bids as inputs and controls the boost vector $z$; it does not redesign advertisers' bidding policies. Our principal contributions are:
\begin{itemize}
  \item \textbf{Quality-aware welfare guarantees and revenue trade-offs.} We develop a capped benchmark-competitive certificate for quality-aware liquid welfare. We derive a welfare lower bound parameterized by the certificate strength, boost cap, and bid cap, explicitly quantifying the shortfall caused by a finite boost cap. A revenue-explicit bound and an impossibility result characterize the welfare--revenue trade-off under our auction and payment rule.
  \item \textbf{Canonical boost design and the price of robustness.} We derive the canonical boost, a closed-form certified rule that maximizes retained revenue among certified implementations of the benchmark allocation whenever it seats every benchmark winner. We quantify its additional protection against rival-bid increases relative to minimal-flip and bound the associated loss in retained revenue.
  \item \textbf{Welfare preservation under bid shocks on AuctionNet.} We evaluate our mechanism on AuctionNet~\cite{su2024auctionnet}, a public benchmark built on Alibaba's real-world production advertising environment. Across ten fixed bidding policies and five episodes, canonical achieves $97.6\%$ mean welfare efficiency under a twofold shock to the strongest rival's bid while the strongest benchmark baselines are $91.1\%$. (Without the twofold shock, they attain near-full efficiency at $99.4\%$). A controlled single-rival stress test isolates the robustness margin, while a boost-cap sweep measures its welfare benefit and payment cost.
\end{itemize}

\subsection{Related work}
\runin{Boosted auctions and auto-bidding.} Boosts have been studied as tools for
revenue optimization and allocation control in sponsored-search and auto-bidding
auctions \cite{edelman2007internet,golrezaei2017boosted,golrezaei2021boosted,
likhodedov2004methods,sandholm2015automated,thompson2013revenue}. Recent auto-bidding analyses study
budget-constrained bidders and robust auction design \cite{deng2021towards,
balseiro2021robust,deng2023autobidding,dobzinski2014efficiency,lu2017liquid,
hao2020dynamic,weinan2016feedback}.
These works generally optimize commercial value or revenue. They do not model a
publisher-side quality externality in the welfare benchmark. Our capped
certificate makes the trade-off between quality-aware welfare and revenue explicit.
Deng et al.~\cite{deng2021towards} prove a benchmark-competitive welfare guarantee
for a two-party (advertiser and platform) auction. Whether that guarantee extends
to our three-party setting, where publisher quality enters both the ranking score
and the welfare benchmark, is not known.
Section~\ref{sec:bound} addresses this question.

\runin{Quality scoring and advertising experience.} Prior work studies quality
scores, click and conversion signals, and post-click user experience
\cite{ghose2009empirical,jain2014evaluating,jansen2017conversion,
lalmas2015promoting,yuan2014analyzing,yoon2010optimal}. Other studies examine
auctions that combine price and quality
\cite{awaya2025quality,camboni2025bidding,huang2019empirical,wang2020quality}.
We use these results to motivate a publisher quality term. Quality remains
external to the advertiser's bid objective, and the platform controls only $z$.

\runin{Learning-based bidding.} AuctionNet and related auto-bidding benchmarks
evaluate offline RL and behavior-cloning policies, including IQL, CQL, BCQ, MOPO,
and COMBO \cite{su2024auctionnet,kostrikov2021offline,kumar2020conservative,
fujimoto2019off,fujimoto2021minimalist,yu2020mopo,yu2021combo,torabi2018behavioral}.
We use these policies as bidder-side inputs and compare boost mechanisms while
holding the bidding strategy fixed. The contribution is the canonical boost
layer, not a new bidder learner.

Section~\ref{sec:model} presents the three-stakeholder online bidding model.
Section~\ref{sec:bound} develops the trade-off theory for welfare and revenue.
Section~\ref{sec:qboost} derives the canonical boost and its price of robustness.
Section~\ref{sec:exp} reports the AuctionNet experiments.

\section{Three-Stakeholder Model}
\label{sec:model}

\begin{figure}
    \centering
    \includegraphics[width=\linewidth]{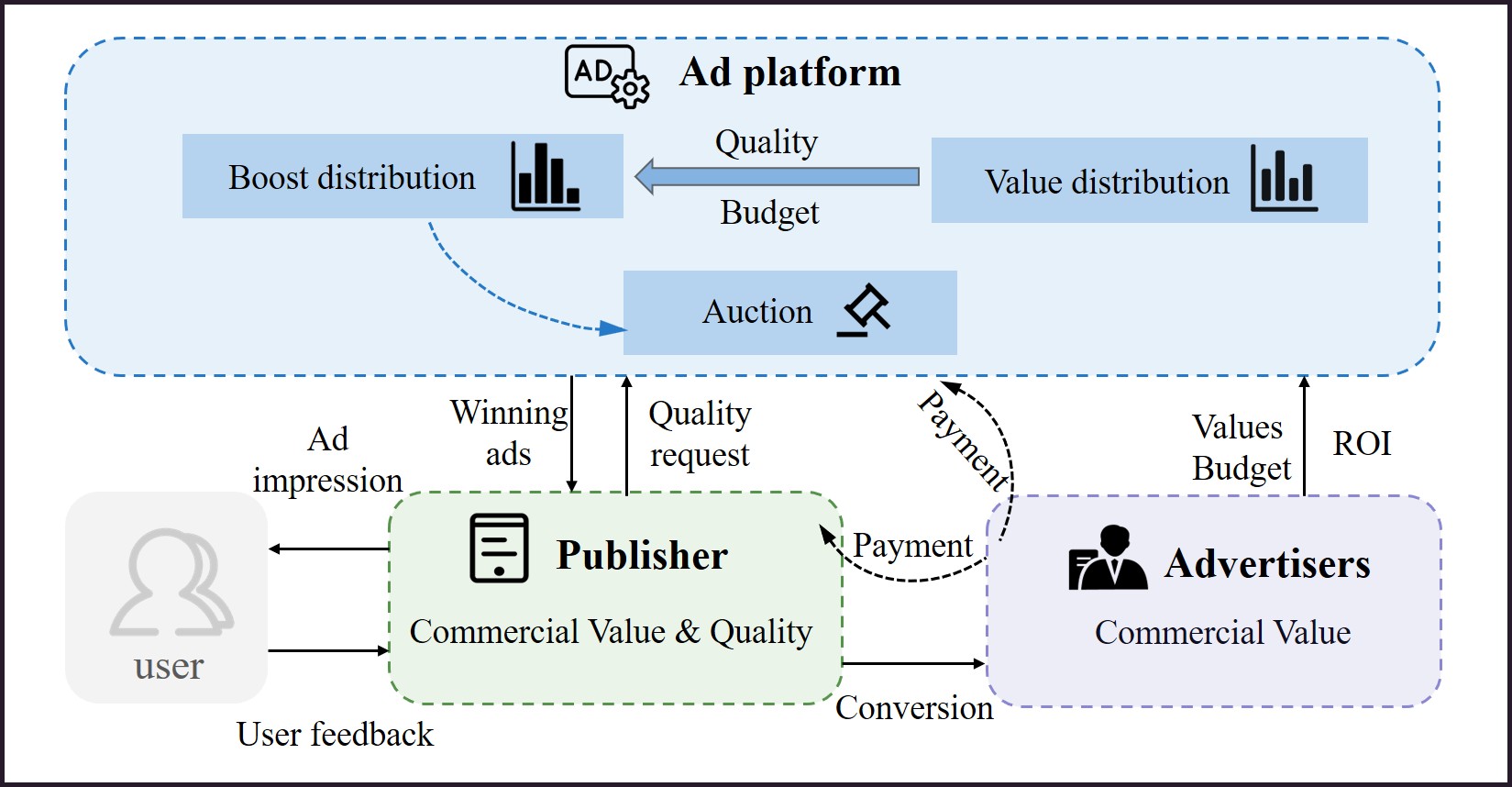}
    \caption{Three-stakeholder online bidding system.}
    \Description{Three-stakeholder online bidding system.}
    \label{fig:system}
\end{figure}

Figure~\ref{fig:system} summarizes the three-stakeholder ecosystem. Advertisers
submit ads, commercial values, budgets, and bids. The publisher supplies
impression opportunities and quality assessments. The platform matches ads to
impressions and runs the auction. Users generate requests and feedback but are
not strategic agents in our model. At each tick, the submitted bids, values,
qualities, and budgets form the auction state. The platform chooses the boost
vector $z$ at that tick, while the scalar certificate parameters are fixed at
design time in Section~\ref{sec:qboost}. The publisher's quality assessment is
also a state variable. The tension is between advertiser value, publisher-side
quality, and platform revenue.
In this paper we study a \emph{single-slot} auction and leave multi-slot settings to future work.

\subsection{Quality and welfare}
We analyze a single bidding tick in which $n$ advertisers compete for $m$
single-slot impression opportunities. For advertiser $i$ on opportunity $j$, let
$v_{i,j}\ge0$ be the commercial value of a conversion and $B_i$ the advertiser's
budget. The \emph{quality value} $q_{i,j}\ge0$ is the publisher's assessment of the
ad, initialized from the ad category and refined over time by user reports. It
is a publisher-side quantity and never enters any advertiser's objective.
Quality estimation in advertising commonly combines click-through, conversion,
landing-page, and post-click signals~\cite{ghose2009empirical,jain2014evaluating,
jansen2017conversion,lalmas2015promoting}. The resulting ad ranking problem is
also connected to pairwise and listwise learning-to-rank methods
\cite{cao2007learning,xia2008listwise}.

To measure the outcome we adopt \emph{liquid welfare}, the standard efficiency
notion for budget-constrained bidders: it credits each advertiser with the smaller
of her budget and the value she receives, and matches the total value attainable
under full budget information~\cite{dobzinski2014efficiency,lu2017liquid,
deng2021towards}. Writing
$x_{i,j}=\mathbf 1\{i=w(j)\}$ and $V_i(x)=\sum_j v_{i,j}x_{i,j}$, the commercial and
quality components are
\[
  \Wel_c(x)=\sum_{i}\min\{B_i,V_i(x)\},
  \qquad
  \Wel_q(x)=\sum_{i,j}q_{i,j}x_{i,j}=\sum_{j}q_{w(j),j},
\]
so the \emph{social welfare} of allocation $x$ is their sum,
\begin{equation}\label{eq:wel}
  \Wel(x)=\Wel_c(x)+\xi\Wel_q(x)
        =\sum_{i}\min\{B_i,V_i(x)\}+\sum_{j}\xi q_{w(j),j}.
\end{equation}
Payments are transfers inside the system, so they cancel out of the welfare sum
and do not appear in \eqref{eq:wel}. The objective is therefore the
planner-weighted sum of advertiser value and publisher experience, with no ad hoc
term~\cite{yoon2010optimal}. The \emph{benchmark} is the welfare-optimal allocation
$O=\argmax_x\Wel(x)$ over $x\in\{0,1\}^{n\times m}$ with $\sum_i x_{i,j}\le1$ for every $j$, and $o(j)$ denotes its winner on impression $j$. We set $\xi=1$ throughout, treating $v$ and $q$ as measured on comparable scales. A platform with a differently scaled quality score absorbs the difference in $\xi$, and every bound below extends verbatim to $\xi\ne1$ by replacing $q$ with $\xi q$. One quality unit is then one unit of welfare, a normalization rather than a monetarily calibrated exchange rate, and the reported trade-offs are conditional on it.

\subsection{Allocation, payment, and advertiser behavior}
The platform ranks ads by a composite \emph{score} that adds the quality value
and a boost $z_{i,j}\in[0,Z_{\max}]$, where $Z_{\max}$ is the boost cap, to the bid $b_{i,j}$. The highest score wins,
\begin{equation}\label{eq:score}
  s_{i,j}=b_{i,j}+q_{i,j}+z_{i,j},\qquad
  w(j)=\argmax_{1\le i\le n} s_{i,j}.
\end{equation}
Exact score ties use the larger boost as a secondary key. For the certified and
minimal-flip rules, we call the benchmark winner $o(j)$ the \emph{anchor}; it is
an explicit final tie key, and other ties use stable index order. Thus the anchor
wins the tie it creates.
The winner pays the second-highest score net of the
quality and boost she
enjoyed,
\begin{equation}\label{eq:pay}
  p_{w(j),j}=\POW{s_{(2),j}-q_{w(j),j}-z_{w(j),j}},
\end{equation}
where $s_{(2),j}$ is the second-highest score and $p_{i,j}=0$ for $i\ne w(j)$. The positive part prevents the
platform from subsidizing a winner. Subtracting $q+z$ keeps the payment at or
below the winner's bid. Rebating $q$ and $z$ is a mechanism-design
choice, not a property of boosted auctions in general. It turns the boost into a
revenue cost and creates the trade-off studied below. The rule extends to oCPX-style
objectives when the platform's ranking score is a linear combination of bid and
quality signals. A full treatment of pCVR/pCTR
estimation, target-CPA constraints, and budget pacing is outside the present
model.

Advertisers choose bids to maximize received commercial value under a budget constraint,
\begin{equation}\label{eq:adv}
  \max_{b_{i,j}}\ \sum_{j}v_{i,j}x_{i,j}
  \quad\text{s.t.}\quad
  \sum_{j}p_{i,j}\le B_i,
\end{equation}
with quality deliberately absent. Advertisers do not internalize it. The theory
conditions on the submitted bids and imposes the $\kappa$-bounded cap stated below.
Because the payment rule gives $p_{i,j}\le b_{i,j}$, this cap keeps the total payment
within a factor $\kappa$ of commercial liquid welfare.

\begin{assumption}[$\kappa$-bounded bids]\label{ass:kappa}
Every advertiser submits a nonnegative bid bounded by a fixed multiple of value,
\[
0\le b_{i,j}\le\kappa\,v_{i,j},\qquad \kappa\ge1.
\]
Together with the budget constraint and the payment rule, this implies
$\Rev(x)\le\kappa\,\Wel_c(x)\le\kappa\,\Wel(x)$.
\end{assumption}

When $\kappa=1$ this also implies value-based
individual rationality ($p\le b\le v$). For $\kappa>1$ it only bounds the payment
by the submitted bid, not by value. Assumption~\ref{ass:kappa} converts the revenue-explicit bound of Corollary~\ref{cor:rev} into a pure-welfare constant. Results that rely on it
are flagged in Section~\ref{sec:bound}. The guaranteed constant weakens
monotonically in $\kappa$. We interpret $\kappa$ as a platform-enforced
bid cap: automated bidders can overbid, so the experiments project every
submitted bid onto the feasible set,
$b_{i,j}\leftarrow\min\{b_{i,j},\kappa\,v_{i,j}\}$ (Sec.~\ref{sec:exp}). All
theory statements apply to the post-projection bids seen by the auction.

Consequently, the platform's control problem is to choose $z$ (subject to the
boost cap and the auction/payment rule) to maximize welfare,
$\max_z \Wel(x(z))$, conditional on the submitted bids. Since every unit of boost
is deducted from a payment, the same choice also changes revenue
$\Rev(x)=\sum_j p_{w(j),j}$. This is the source of the trade-off between welfare and revenue
studied next. Advertiser bidding is therefore an input to the platform's boost
design problem, not an additional decision variable in Sections~\ref{sec:bound}
and~\ref{sec:qboost}. The model conditions on the submitted bids. Strategic bid
shading is outside the model and is discussed in Section~\ref{sec:qboost}.

\section{Welfare Guarantees under Revenue Costs}
\label{sec:bound}
Every unit of boost added to a winner's score is deducted from her payment,
Eq.~\eqref{eq:pay}. Using boosts to favor the welfare-optimal allocation can
therefore reduce revenue, as the following example shows.

\begin{example}[Welfare versus revenue]\label{ex:tradeoff}
Consider one impression with a value-rich advertiser~1
$(v_1,b_1,q_1,B_1)=(8,8,0,\infty)$ and a quality-rich advertiser~2
$(v_2,b_2,q_2,B_2)=(7,3.5,4,7)$. Without a boost, advertiser~1 wins because
$8>7.5$, yielding $(\Wel,\Rev)=(8,7.5)$. The welfare optimum seats
advertiser~2, whose welfare is $7+4=11$. A minimal boost $z_2=0.5$ makes its
score tie at $8$ and gives the tie to the boosted bidder. Since quality and
boost are deducted from payment, the resulting outcome is
\[
  \underbrace{(\Wel,\Rev)=(8,\,7.5)}_{\text{no boost}}
  \qquad\text{versus}\qquad
  \underbrace{(\Wel,\Rev)=(11,\,3.5)}_{\text{boost } z_2=0.5},
\]
Thus, three welfare units cost four revenue units. Boosting beyond the winning
margin changes neither allocation nor welfare.
\end{example}

The example captures a common pattern. The welfare-optimal winner may have high
quality but a low bid, so the boost that lets it win reduces the platform's
payment. We therefore construct a certificate set of feasible boosts and
quantify the welfare guarantee and the lower-bound envelope relating welfare to revenue
that every boost in the set satisfies. Section~\ref{sec:qboost} studies the
optimization over this certified set.

We measure guarantees against the welfare-optimal allocation
$O=\argmax_x\Wel(x)$, whose winner on impression $j$ we call the \emph{benchmark
winner} $o(j)$, and we let the \emph{efficiency} of a single tick be
$\mathrm{Eff}(x)=\Wel(x)/\Wel(O)$. We introduce a free, nonnegative
\emph{margin coefficient} $c$. It scales the benchmark winner's
commercial and quality values $v_{o(j),j}+q_{o(j),j}$ in the required boost
difference. The additional term $\POW{v_{o(j),j}-b_{o(j),j}}$ compensates for bid
shading.
A larger $c$ requires a larger boost difference until the cap $Z_{\max}$ binds.
This defines a family of certificates. Theorem~\ref{thm:main} establishes its
quality-aware welfare guarantee rather than assuming a classical advertiser-only
bound.
\begin{definition}[Capped benchmark-competitive boost]\label{def:bc}
Fix $c\ge0$ and the boost cap $Z_{\max}$, and set
$M_j=c(v_{o(j),j}+q_{o(j),j})+\POW{v_{o(j),j}-b_{o(j),j}}$ and
$m_j=\min\{M_j,Z_{\max}\}$. A boost $z$ is \emph{capped benchmark-competitive},
written $z\in\Zbc$, iff
\[
\begin{aligned}
  &0\le z_{i,j}\le Z_{\max}
  \quad\text{and}\quad
  z_{o(j),j}-z_{i,j}\ \ge\ m_j\\
  &\qquad\qquad\qquad\qquad\qquad\qquad
  \forall\, j,\ \forall\, i\ne o(j).
\end{aligned}
\]
\end{definition}
When $M_j\le Z_{\max}$ this is a budget-aware, quality-aware strengthening of the
benchmark-competitive requirement of~\cite{deng2021towards}: the margin is scaled by
$v_{o(j),j}+q_{o(j),j}$ rather than by value alone, and the bid discount
$\POW{v_{o(j),j}-b_{o(j),j}}$ compensates for shading. When $M_j>Z_{\max}$ the
required margin cannot be met, and the shortfall becomes the penalty term discussed further in Theorem~\ref{thm:main}. Because $m_j\le Z_{\max}$, the boost
$z_{o(j),j}=m_j$ with $z_{i,j}=0$ for $i\ne o(j)$ satisfies both constraints, so
$\Zbc$ is non-empty and the certificate is never vacuous.

The next lemma explains why this class trades welfare against revenue: a boost, and
likewise a quality score, is paid for out of the winner's own payment.
\begin{lemma}[Boost transfers revenue]\label{lem:couple}
For the winner $w=w(j)$, $p_{w,j}\le b_{w,j}$. Up to the clamp, every unit of
boost or quality granted to the winner is a unit of forgone payment.
\end{lemma}

We can now state the certificate guarantee. If the budget meets every required
margin, the unpenalized bound $\frac{c+1}{c+1+\kappa}\Wel(O)$ is recovered.
Otherwise the guarantee shrinks with the total shortfall.
\begin{theorem}[Bounded-boost welfare lower bound]\label{thm:main}
Fix any $c\ge0$. Under Assumption~\ref{ass:kappa} with parameter $\kappa$, for any $z\in\Zbc$, let
$A=\{j:w(j)=o(j)\}$ and
$\delta_j=\POW{(v_{o(j),j}+q_{o(j),j})-\frac{b_{o(j),j}+q_{o(j),j}+m_j}{c+1}}$.
Then
\begin{align}\label{eq:main}
  \Wel(x)\ &\ge\ \frac{c+1}{c+1+\kappa}\Big(\Wel(O)-\sum_{j\notin A}\delta_j\Big)
  \ \nonumber \\
  &\ge\ \frac{c+1}{c+1+\kappa}\,\Wel(O)
  -\frac{c+2}{c+1+\kappa}\,\Lambda(Z_{\max}),
\end{align}
where
$\Lambda(Z_{\max})=\frac{1}{c+2}\sum_{j:\,M_j>Z_{\max}}
\POW{(c+1)(v_{o,j}+q_{o,j})-b_{o,j}-q_{o,j}-Z_{\max}}$.
Equivalently $\Wel(x)\ge\rho_\kappa(Z_{\max})\,\Wel(O)$ with
\[
  \rho_\kappa(Z_{\max})
  =\frac{c+1}{c+1+\kappa}-\frac{c+2}{c+1+\kappa}\frac{\Lambda(Z_{\max})}{\Wel(O)},
\]
the ratio $\rho_\kappa$ is non-decreasing in $Z_{\max}$, and
$\rho_\kappa=\frac{c+1}{c+1+\kappa}$ once $Z_{\max}\ge\max_j M_j$.
\end{theorem}


The proof is deferred to Appendix~\ref{app:thm1}. The penalty $\Lambda$ is the
accumulated form of the loss seen in Example~\ref{ex:tradeoff}: it vanishes only
when the budget is large enough to boost every benchmark winner by its full margin.
The statement is per tick and conditional on the state $(v,q,b,B)$. It assumes
only the $\kappa$-bounded bid cap, not truthful bidding. Section~\ref{sec:bound}
supplies the certified feasible domain and its guarantee. Section~\ref{sec:qboost}
shows that the same domain singles out a revenue-maximal implementation.

Because every boost is paid for out of the winner's own payment
(Lemma~\ref{lem:couple}), the proof in Appendix~\ref{app:thm1} actually yields a
sharper, \emph{revenue-explicit} guarantee: the welfare loss from the optimum is
bounded by the realized revenue, scaled by $1/(c+1)$, plus the boost-budget
penalty.
\begin{corollary}[Revenue-parameterized welfare bound]\label{cor:rev}
For every $z\in\Zbc$,
\begin{equation}\label{eq:correv}
  \Wel(x)\ \ge\ \Wel(O)-\!\!\sum_{j\notin A}\!\delta_j-\frac{\Rev(x)}{c+1}
  \ \ge\ \Wel(O)-D(Z_{\max})-\frac{\Rev(x)}{c+1},
\end{equation}
where $D(Z_{\max})=\sum_{j:\,M_j>Z_{\max}}\delta_j=\frac{c+2}{c+1}\Lambda(Z_{\max})$.
Equivalently, $\Wel(x)+\frac{1}{c+1}\Rev(x)\ \ge\ \Wel(O)-\sum_{j\notin A}\delta_j$.
The revenue-explicit form \eqref{eq:correv} is \emph{assumption-free}: it does not
use Assumption~\ref{ass:kappa}. Under that assumption $\Rev(x)\le\kappa\,\Wel(x)$
(Appendix~\ref{app:thm1}), so \eqref{eq:correv} also implies the constant form
in Theorem~\ref{thm:main}. It is the sharper, revenue-aware version of the bound:
a mechanism collecting revenue $\Rev$ has worst-case welfare at least
$\Wel(O)-D(Z_{\max})-\Rev/(c+1)$, so its welfare shortfall cannot exceed the
revenue, converted at rate $1/(c+1)$.
\end{corollary}

\begin{remark}[Lower-bound envelope for welfare and revenue]\label{rem:envelope}
Under Assumption~\ref{ass:kappa}, combining Corollary~\ref{cor:rev} with the
constant bound of Theorem~\ref{thm:main} yields the envelope
\[
  \Wel(x)\ \ge\ \max\Big\{\ \tfrac{c+1}{c+1+\kappa}\Big(\Wel(O)-\!\!\sum_{j\notin A}\!\delta_j\Big),\ \
  \Wel(O)-\!\!\sum_{j\notin A}\!\delta_j-\tfrac{\Rev(x)}{c+1}\ \Big\}.
\]
Up to the knee
$\Rev^\ast=\tfrac{\kappa(c+1)}{c+1+\kappa}(\Wel(O)-\sum_{j\notin A}\delta_j)$ the
guarantee decreases linearly in the realized revenue at rate $1/(c+1)$. Beyond
$\Rev^\ast$ the certified floor saturates at the budget-parameterized bound
$\rho_\kappa(Z_{\max})\Wel(O)$. Each unit of revenue retained up to $\Rev^\ast$
may thus cost up to $1/(c+1)$ units of worst-case welfare. The saturation beyond
the knee is a statement about the bound, not about the realized welfare.
\end{remark}

The envelope above quantifies the cost of revenue \emph{inside} the certificate.
It leaves open whether a different boost, or a different mechanism, could avoid
it. The next result rules out the first possibility: for any target $\delta$ and any
$\varepsilon>0$ there is an instance on which every boost under the auction and
payment rule of Section~\ref{sec:model} that attains at least $(1-\delta)$
welfare collects at most $\varepsilon\,\Wel(O)$ in revenue. The trade-off is
therefore not an artifact of the certificate class. Whether a different mechanism
could do better is outside the present model.
\begin{corollary}[Unavoidable trade-off]\label{cor:imposs}
Under Assumption~\ref{ass:kappa}, for every $\delta\in(0,\tfrac12)$, every
$\kappa\ge1$, and every $\varepsilon>0$ there is an instance in which
$\Wel(x)\ge(1-\delta)\Wel(O)$ forces $\Rev(x)\le\varepsilon\,\Wel(O)$, for every
boost $z$ under the auction and payment rule of Section~\ref{sec:model}.
\end{corollary}

The construction separates an ad's welfare contribution from its ability to
pay. A high-quality ad can be the welfare-optimal choice even when its commercial
value is arbitrarily small. Approaching the welfare optimum then requires
selecting this ad. Under our payment rule, however, its payment cannot exceed
its bid, and the bid cap limits that bid to $\kappa$ times commercial value.
Publisher quality raises welfare without raising this payment ceiling. The
argument applies to every boost vector, including those outside the certified
set.

\runin{A computable reference allocation.} Theorem~\ref{thm:main}
requires only a reference winner lifted above every rival by the margin $M_j$.
It never uses the optimality of $O$. Thus, when $O$ is unavailable at scale, any
reference allocation $R$ with winner $o_R(j)$ yields the same guarantee with
$\Wel(O)$ replaced by $\Wel(R)$.

\begin{corollary}[Computable reference allocation]\label{cor:ref}
Under Assumption~\ref{ass:kappa}, for any allocation $R$ with per-impression
winner $o_R(j)$, let $z$ be capped benchmark-competitive with respect to $R$
(Definition~\ref{def:bc} with $o_R$ in place of $o$ and
$M_j=c(v_{o_R(j),j}+q_{o_R(j),j})+\POW{v_{o_R(j),j}-b_{o_R(j),j}}$). Then
\[
  \Wel(x)\ \ge\ \frac{c+1}{c+1+\kappa}\Wel(R)
  -\frac{c+2}{c+1+\kappa}\Lambda_R(Z_{\max}),
\]
where $\delta_j^R$ and $\Lambda_R$ are defined as in Theorem~\ref{thm:main} with
$R$ in place of $O$. Taking $o_R(j)=\argmax_i(v_{i,j}+q_{i,j})$ makes $R$
computable in $\mathcal{O}(nm)$ time. Since $R$ is feasible, $\Wel(R)\le\Wel(O)$, so this
is a scalable fallback, not a stronger bound. The main formulation and reported
efficiency remain relative to $\Wel(O)$ whenever that benchmark is computed.
\end{corollary}

If a reference algorithm guarantees \(\mathrm{Wel}(R)\ge\gamma\,\mathrm{Wel}(O)\), where \(0<\gamma\le1\), Corollary 3 gives
\[
\mathrm{Wel}(x)\ge
\gamma\frac{c+1}{c+1+\kappa}\mathrm{Wel}(O)
-\frac{c+2}{c+1+\kappa}\Lambda_R(Z_{\max}).
\]
When the cap covers all reference margins, the resulting approximation ratio is \(\gamma(c+1)/(c+1+\kappa)\). The reference approximation scales the certificate's constant, while the boost cap contributes the penalty \(\Lambda_R\). A heuristic reference without a proved \(\gamma\) still supports the comparison to \(\mathrm{Wel}(R)\), but does not by itself establish a constant approximation to optimal welfare.

\begin{remark}[Relation to the classical benchmark-competitive bound]\label{rem:deng}
Under Assumption~\ref{ass:kappa}, as $Z_{\max}\to\infty$, $\Lambda\to0$ and Theorem~\ref{thm:main} reduces to $\Wel(x)\ge\frac{c+1}{c+1+\kappa}\Wel(O)$. When $\kappa=1$, the coefficient coincides with the $(c+1)/(c+2)$ bound of \cite{deng2021towards}; the two results differ: their welfare counts advertiser value only, their margin is $c\,v_o$, and they impose no bid cap. Our welfare includes publisher quality, our margin is $c(v_o+q_o)+\POW{v_o-b_o}$, and bids are bounded by $\kappa v$. Because quality lies outside the budget-capped value $\min\{B_i,V_i(x)\}$, it cannot be absorbed into $v$ by a change of variables, so the quality-aware bound is proved independently in Appendix~\ref{app:thm1}. The special case with no quality term, $\kappa=1$, and $Z_{\max}\ge\max_j M_j$ recovers their single-slot, unit-bid result.
\end{remark}

\section{Certified Boost Design}
\label{sec:qboost}

Section~\ref{sec:bound} delivers a \emph{certificate} rather than a single
mechanism: every boost in the capped benchmark-competitive set $\Zbc$ attains
welfare at least $\rho_\kappa(Z_{\max})\Wel(O)$ against the welfare-optimal
allocation $O=\argmax_x\Wel(x)$. This section shows that the same set also
\emph{determines} the boost. On every impression where its margin seats the
benchmark winner, the canonical element maximizes the winner's payment among all
certified boosts that also seat her. When this holds on every impression, it
maximizes the platform's retained revenue $\Rev(x)$ among all
certified implementations of the benchmark allocation. It can be written in
closed form.
\subsection{The design problem}
Section~\ref{sec:bound} fixes the welfare side of the design: every
$z\in\Zbc$ attains the guarantee of Theorem~\ref{thm:main}. The remaining
question is how to implement the benchmark allocation at the lowest revenue
cost. Since payments are transfers, implementing $O$ attains $\Wel(O)$, the
welfare optimum. We therefore use a lexicographic objective: first target $O$,
then minimize the revenue cost among certified implementations of $O$.
Formally, the platform solves
\begin{equation}\label{eq:design}
  \max_{z}\ \Rev(x(z))
  \quad\text{s.t.}\quad
  z\in\Zbc,\quad w(j)=o(j)\ \ \forall j,
\end{equation}
where $o(j)$ is the per-impression winner of $O$,
and $\Zbc$ is the set of Definition~\ref{def:bc}. Because both the certificate
constraints and the payments are per impression, \eqref{eq:design} separates
across impressions. The certificate constraint is exactly the sufficiency
condition used by Theorem~\ref{thm:main} (or Corollary~\ref{cor:ref} if using $\Wel(R)$), so no feasible $z$ loses the
guarantee. Proposition~\ref{prop:canonical} solves \eqref{eq:design} on every impression
where the canonical boost seats $o$. On the remaining impressions, the
certificate may still admit another implementation of $O$, but we do not
characterize the revenue-maximal one. If no certified boost implements $O$ on
such an impression, the per-impression problem is infeasible and we fall back
to the guarantee of Theorem~\ref{thm:main} rather than exact optimality.
Section~\ref{sec:exp} reports the realized ratio and the cap-binding rate.

\subsection{The canonical boost}
Consider the boost that lifts the benchmark
winner by exactly the certificate margin $m_j$ and leaves every rival unboosted,
\begin{equation}\label{eq:canonical}
  z^{c}_{o(j),j}=m_j,
  \qquad
  z^{c}_{i,j}=0\quad (i\ne o(j)).
\end{equation}
This boost is feasible by construction. The next result shows that it is also
revenue-maximal among the certified boosts that implement the benchmark allocation.

\begin{proposition}[Canonical boost]\label{prop:canonical}
Fix an impression $j$, let $o=o(j)$, and define
$z^{c}$ by \eqref{eq:canonical}. Then $z^{c}\in\Zbc$. If $z^{c}$ seats $o$, then
for every $z\in\Zbc$ that also seats $o$,
\begin{equation}\label{eq:canon-pay}
  p_{o,j}(z)\ \le\ p_{o,j}(z^{c})
  =\Big[\max_{i\ne o}\,(b_{i,j}+q_{i,j})-q_{o,j}-m_j\Big]^{+}.
\end{equation}
\end{proposition}

\begin{proof}
Feasibility is immediate: $0\le z^{c}\le Z_{\max}$ and
$z^{c}_{o,j}-z^{c}_{i,j}=m_j$ for every $i\ne o$. For any feasible $z$ that
seats $o$, the second-highest score satisfies
\[
\begin{aligned}
  s_{(2),j}
  &=\max_{i\ne o}\,(b_{i,j}+q_{i,j}+z_{i,j})
  \le \max_{i\ne o}\,(b_{i,j}+q_{i,j})+\max_{i\ne o}z_{i,j}\\
  &\le \max_{i\ne o}\,(b_{i,j}+q_{i,j})+z_{o,j}-m_j,
\end{aligned}
\]
where the last step uses the certificate $z_{i,j}\le z_{o,j}-m_j$. Substituting
into $p_{o,j}=[s_{(2),j}-q_{o,j}-z_{o,j}]^{+}$ yields
\eqref{eq:canon-pay}. The canonical boost attains the bound because its second
score is exactly $\max_{i\ne o}(b_{i,j}+q_{i,j})$.
\end{proof}

Proposition~\ref{prop:canonical} shows that the certificate fixes the benchmark
winner's lead at $m_j$, and any additional boost only transfers payment away from
the platform. We call $z^{c}$ \emph{canonical} because, among the revenue-maximal
certified implementations of the benchmark allocation, it is the unique
representative with every rival unboosted. Any common shift of all boosts
preserves the payment but wastes budget. When $z^{c}$
implements $O$, summing \eqref{eq:canon-pay} over impressions gives
$\Rev(x(z))\le\Rev(x(z^{c}))$ for every $z\in\Zbc$ that also implements $O$.
The per-impression statement is therefore also a statement about total retained
revenue. The certificate therefore selects the
revenue-maximal way to implement the benchmark allocation. The platform's
remaining decisions are the three scalar parameters of Section~\ref{sec:levers}. The condition ``if
$z^{c}$ seats $o$'' is not automatic: when a rival's bid exceeds the margin, the
canonical rule can fail to implement the benchmark winner, and the payment bound
then applies to any certified boost that does implement it. Section~\ref{sec:exp}
reports the realized welfare ratio and the cap-binding rate used to interpret this
condition.

\subsection{The price of robustness}
The canonical boost is not the cheapest way to seat the benchmark winner. The
minimal-flip boost uses only the margin needed under the observed bids,
\begin{equation}\label{eq:flip}
\begin{aligned}
  m^{f}_j&=\max\Big\{0,\ \max_{i\ne o}\,(b_{i,j}+q_{i,j})-(b_{o,j}+q_{o,j})\Big\},\\
  z^{f}_{o(j),j}&=\min\{m^{f}_j,Z_{\max}\},
\end{aligned}
\end{equation}
and leaves rivals unboosted. Whenever $m^{f}_j\le Z_{\max}$, it implements the
same allocation with a smaller boost, so it retains more revenue when
$m^{f}_j<m_j$. If $m^{f}_j>Z_{\max}$, the cap prevents it from seating the
benchmark winner. When $m^{f}_j\le m_j$, the difference $m_j-m^{f}_j$ is the
certificate's robustness margin. Minimal-flip is not a certified boost:
$m^{f}_j$ can be strictly smaller than $m_j$, so $z^{f}$ need not belong to
$\Zbc$. Proposition~\ref{prop:canonical}'s revenue-maximality therefore does
not apply to it. Exact score ties are resolved in favor of the boosted anchor
$o(j)$. This is a
tie-breaking rule rather than an economically meaningful boost. The
implementation uses a $10^{-9}$ numerical epsilon for this purpose, far below
any economic effect, and the auction uses the same rule as a secondary key
(larger boost first), so the cap is never violated and the formulas below hold
up to this negligible epsilon.

\begin{proposition}[Robustness margin]\label{prop:robust}
Let $S_j=\max_{i\ne o}\,(b_{i,j}+q_{i,j})$ and
$m^{f}_j=\max\{0,\,S_j-(b_{o,j}+q_{o,j})\}$, and suppose $m^{f}_j\le m_j$ and
$m^{f}_j\le Z_{\max}$. If every rival's score increases by at most $\Delta_j$, with
$\Delta_j\le m_j-m^{f}_j$, while the anchor's bid is unchanged, the canonical
boost still seats $o(j)$. Under the minimal-flip boost the residual lead is $0$
when $S_j\ge b_{o,j}+q_{o,j}$, so any $\Delta_j>0$ can overturn the allocation.
\end{proposition}

\begin{proof}
If $S_j\ge b_{o,j}+q_{o,j}$, then $m^{f}_j=S_j-(b_{o,j}+q_{o,j})$ and the
residual lead of $z^{c}$ is $(b_{o,j}+q_{o,j}+m_j)-S_j=m_j-m^{f}_j$. After a
rival score increase $\Delta_j$ the lead is $m_j-m^{f}_j-\Delta_j\ge0$ by the
stated condition. At equality the tie goes to $o$. Under $z^{f}$ the lead is
$(b_{o,j}+q_{o,j}+m^{f}_j)-S_j=0$, so any $\Delta_j>0$ overturns it. If
$S_j<b_{o,j}+q_{o,j}$, then $m^{f}_j=0$, both rules already seat $o$, and the
certified lead is larger by $m_j$. The same argument applies with the existing
lead added.
\end{proof}

The extra margin $m_j-m^{f}_j$ is thus a \emph{robustness budget} whenever
$m^{f}_j\le m_j$. Directly from the payment formula (and $m^{f}_j\le m_j$),
\[
  0\ \le\ p_{o,j}(z^{f})-p_{o,j}(z^{c})\ \le\ m_j-m^{f}_j,
\]
with equality when the canonical payment is not clamped (which implies the
minimal-flip payment is not clamped), i.e.\ when
$S_j-q_{o,j}-m_j\ge0$. The payment cost of the extra margin is at most the
margin itself. The experiments measure this trade-off. Minimal-flip retains
more revenue on a static, perfectly observed bid stream. When bids drift
between design and execution, the certificate provides protection that
minimal-flip lacks.

\begin{remark}[Bid shading]\label{rem:bidshading}
The term $\POW{v_o-b_o}$ makes the anchor's score
$b_o+q_o+m_j=c(v_o+q_o)+v_o+q_o$ independent of $b_o$ for $b_o\le v_o$, while
the payment decreases by the same amount as $b_o$. A bidder who shades toward zero
therefore keeps the allocation and welfare but can drive the payment to zero.
The mechanism is not strategy-proof, and the fixed-bid assumption is essential.
For example, with $v_o=7$, $q_o=1$, $b_o=7$ and a rival with $v_r=7.9$ and
score $19.75$, the payment is $2.75$. At $b_o=0$ the margin rises from $16$ to
$23$ and the payment
falls to $0$, while the anchor still wins and welfare is unchanged. Outside the
present model, removing the $\POW{v_o-b_o}$ term, using a bid-independent margin,
or imposing a reserve/minimum payment are natural remedies. Equilibrium analysis
is left to future work.
\end{remark}

\subsection{Three parameters and the operating point}
\label{sec:levers}
The certificate has three scalar parameters: the strength $c$,
the boost cap $Z_{\max}$, and the bid cap $\kappa$.
A larger $c$ demands a larger margin and hence a larger boost. The budget
$Z_{\max}$ caps every boost and controls the penalty
$\Lambda(Z_{\max})$. Once $Z_{\max}\ge\max_j M_j$ the certificate is met exactly,
$\Lambda=0$, and the guarantee is the clean
$\frac{c+1}{c+1+\kappa}\Wel(O)$. The bid cap $\kappa$ bounds overbidding and
enters the same constant. A looser cap weakens the guarantee.

The budget also determines the \emph{regime}. When $Z_{\max}<\max_j M_j$ the cap
binds on most impressions, $m_j=Z_{\max}$, and the canonical rule degenerates to
``max out the benchmark winner.'' The guarantee carries the penalty $\Lambda$.
When $Z_{\max}\ge\max_j M_j$ the full margin is available and the guarantee
reaches its constant form. The experiments report the binding fraction
$\Pr[M_j>Z_{\max}]$ alongside $\rho_\kappa(Z_{\max})$, because a guarantee quoted
without its regime is not interpretable. In particular, a cap of the form
$Z_{\max}=\alpha\max_{i,j}(v_{i,j}+q_{i,j})$ with small $\alpha$ is almost always binding.
The certificate is active once $Z_{\max}\ge\max_j M_j$, which requires $\alpha$
to exceed $c$ times the ratio of the anchor's $(v_{i,j}+q_{i,j})$ to the tick
maximum (and also covers the anchor's $\POW{v_o-b_o}$ term). In our configuration
the cap-binding fraction drops from $0.89$ at $\alpha=1$ to $0.04$ at
$\alpha=2$. The cap sweep in Section~\ref{sec:exp} traces this transition and
its revenue cost.

\section{Experiments}
\label{sec:exp}
\runin{Single-slot replay setting.} Experiments use the same single-slot,
fixed-bid auction with a frozen quality stream. Multi-slot positions and
endogenous bidder or quality adaptation are outside scope.

\begin{table*}[t]
\centering
\footnotesize
\setlength{\tabcolsep}{6pt}
\caption{Welfare and robustness by mechanism. (a) Episode-501 summary
over ten bidding policies at the main operating point. (b) Per-policy mean
2$\times$-shock welfare over episodes 501--505.}
\label{tab:main}
\renewcommand{\arraystretch}{1.2}
\noindent\textbf{(a) Mechanism summary (episode 501).}\\[2pt]
\begin{tabular*}{\textwidth}{@{\extracolsep{\fill}}l cccc c@{}}
\toprule
\rowcolor{headgray}
\textbf{Mechanism} & \textbf{2$\times$ shock} & \textbf{Clean Eff.} & \shortstack{\textbf{Welfare improvement}\\(vs.\ no boost, pp)} & \textbf{5-ep.\ shock} & \textbf{Worst policy} \\
\midrule
\textbf{Canonical boost (proposed)} & \bfirst{98.8\%}{} & \usecond{99.5\%} & \bfirst{+9.8}{} & \bfirst{97.63\%\,$\pm$\,2.24\%}{} & \bfirst{94.9\% (Mix)}{} \\
c-competitive ($c=1.5$) & 91.4\% & 91.4\% & +2.4 & \usecond{91.12\%\,$\pm$\,0.85\%} & \usecond{90.4\% (Mix)} \\
Minimal-flip & \usecond{92.2\%} & \bfirst{99.6\%}{} & \usecond{+3.2} & 91.12\%\,$\pm$\,1.81\% & 84.3\% (Combo) \\
Uniform & 89.4\% & 95.0\% & +0.4 & 88.64\%\,$\pm$\,1.63\% & 79.7\% (CQL) \\
BSP-AM & 87.4\% & 90.5\% & $-$1.6 & 84.68\%\,$\pm$\,5.71\% & 82.4\% (CQL) \\
Myerson (unclipped)$^{\dagger}$ & 91.9\% & 94.5\% & +2.9 & 90.15\%\,$\pm$\,3.33\% & 88.5\% (CQL) \\
No boost (= clipped Myerson) & 89.0\% & 92.6\% & 0.0 & 86.19\%\,$\pm$\,5.50\% & 77.9\% (CQL) \\
\bottomrule
\end{tabular*}

\vspace{6pt}
\noindent\textbf{(b) Per-policy mean 2$\times$-shock welfare over episodes 501--505: CNY and efficiency.}\\[2pt]
{\scriptsize
\setlength{\tabcolsep}{2pt}
\begin{tabular*}{\textwidth}{@{\extracolsep{\fill}}l ccccccc r@{}}
\toprule
\rowcolor{headgray}
\textbf{Policy} & \textbf{Canonical} & \textbf{c-competitive} & \textbf{Minimal-flip} & \textbf{Uniform} & \textbf{BSP-AM} & \textbf{No boost} & \textbf{Myerson (unclipped)}$^{\dagger}$ & \textbf{Gap vs. best (CNY, pp)} \\
\midrule
PID & \bfirst{541.4 (99.76\%)}{} & 504.0 (92.43\%) & 516.3 (94.86\%) & 512.8 (94.14\%) & 496.1 (90.91\%) & 510.0 (93.60\%) & \usecond{516.7 (94.92\%)} & \gain{+24.7 (+4.84\,pp)} \\
OnlineLP & \bfirst{526.1 (97.74\%)}{} & 495.6 (91.79\%) & 497.2 (92.10\%) & \usecond{497.5 (92.16\%)} & 463.3 (85.51\%) & 497.2 (92.10\%) & 497.4 (92.13\%) & \gain{+28.6 (+5.58\,pp)} \\
BC & \bfirst{542.5 (99.99\%)}{} & 504.3 (92.50\%) & \usecond{540.2 (99.54\%)} & 528.3 (97.21\%) & 428.4 (83.62\%) & 453.3 (88.45\%) & 507.7 (94.95\%) & \gain{+2.3 (+0.45\,pp)} \\
Combo & \bfirst{512.8 (95.99\%)}{} & \usecond{482.4 (90.02\%)} & 440.7 (81.88\%) & 446.5 (83.01\%) & 430.1 (81.21\%) & 437.0 (81.19\%) & 462.7 (86.21\%) & \gain{+30.4 (+5.97\,pp)} \\
BCQ & \bfirst{507.9 (95.01\%)}{} & \usecond{474.7 (88.53\%)} & 446.4 (83.06\%) & 446.9 (83.15\%) & 469.1 (86.63\%) & 446.4 (83.06\%) & 461.7 (86.02\%) & \gain{+33.2 (+6.48\,pp)} \\
CQL & \bfirst{537.0 (98.89\%)}{} & \usecond{505.7 (92.76\%)} & 478.0 (89.11\%) & 432.6 (80.52\%) & 443.3 (81.56\%) & 415.9 (77.12\%) & 466.5 (87.09\%) & \gain{+31.3 (+6.13\,pp)} \\
TD3BC & \bfirst{542.4 (99.96\%)}{} & 504.3 (92.51\%) & \usecond{537.5 (99.02\%)} & 523.1 (96.19\%) & 425.2 (83.02\%) & 447.5 (87.31\%) & 504.1 (94.24\%) & \gain{+4.9 (+0.94\,pp)} \\
IQL & \bfirst{542.2 (99.93\%)}{} & 504.9 (92.63\%) & \usecond{516.4 (95.83\%)} & 484.6 (90.42\%) & 471.3 (87.84\%) & 481.1 (89.76\%) & 489.4 (91.45\%) & \gain{+25.8 (+4.10\,pp)} \\
Mopo & \bfirst{509.1 (95.26\%)}{} & \usecond{474.1 (88.43\%)} & 446.7 (83.10\%) & 447.4 (83.23\%) & 433.6 (82.24\%) & 446.8 (83.11\%) & 461.7 (85.99\%) & \gain{+35.0 (+6.83\,pp)} \\
Mix & \bfirst{505.7 (93.75\%)}{} & 484.5 (89.64\%) & \usecond{500.3 (92.71\%)} & 468.3 (86.42\%) & 457.4 (84.25\%) & 467.4 (86.21\%) & 478.6 (88.46\%) & \gain{+5.4 (+1.04\,pp)} \\
\textbf{Mean of 10} & \bfirst{526.7 (97.63\%)}{} & \usecond{493.5 (91.12\%)} & 492.0 (91.12\%) & 478.8 (88.64\%) & 451.8 (84.68\%) & 460.2 (86.19\%) & 484.6 (90.15\%) & \gain{+33.2 (+6.51\,pp)} \\
\bottomrule
\end{tabular*}
}

\vspace{2pt}
{\footnotesize\noindent $^{\dagger}$\,Myerson (unclipped) allows negative boosts and therefore ranges over a superset of the
nonnegative-boost domain studied here.\par}
\end{table*}

The experiments ask whether the platform can preserve quality-aware welfare
when bids change between mechanism design and execution, and what payment cost
this robustness requires. We first measure the effect of bid drift, then isolate
the mechanism-level source, quantify the payment cost, and check stability.

\subsection{Experimental setup}
\label{sec:exp_setup}
\runin{Data and quality.} We use AuctionNet~\cite{su2024auctionnet}, a public
benchmark built from Alibaba's real-world production advertising environment,
with 48 advertisers and 48 ticks per episode. The environment's quality combines a seed-generated
ad-quality term with a winner-dependent bidder-quality term. We freeze the latter
at zero, so every mechanism sees the same $q$ and the same benchmark.

\runin{Policies and mechanisms.} We compare ten bidding policies (PID, IQL,
TD3BC, OnlineLP, CQL, BC, BCQ, Mopo, Combo, Mix) and seven mechanisms: the
canonical boost, the c-competitive boost~\cite{deng2021towards}, minimal-flip,
a uniform value-proportional boost,
BSP-AM~\cite{golrezaei2021boosted}, unclipped
Myerson, and no boost (= clipped Myerson). Canonical is the only certified
mechanism. The rest are non-certified baselines. In the single slot, the
c-competitive boost uses $c=1.5$ (best parameter in \cite{deng2021towards}).

\runin{Benchmark and reference points.} The benchmark is the
welfare-optimal allocation $O$
under the frozen quality stream, with $\Wel(O)=458.48$ CNY on episode 501. It
is the common denominator for all mechanisms and shocks. In our environment
budgets are slack, so the optimal welfare is easy to compute (non-slack: see
Corollary~\ref{cor:ref}). Due to the page limits, settings and baseline details are placed in Appendix~\ref{app:extra}.

\subsection{Headline result: welfare preservation and shock robustness}
\label{sec:exp_headline}
On the clean fixed-bid replay, canonical and minimal-flip are both near the
welfare ceiling, with mean efficiencies of $99.40\%$ and $99.42\%$ over five
episodes, leaving little separation between them. We therefore use the twofold
stale-bid shock to expose their different robustness to this targeted bid drift.
The result is specific to this perturbation and does not establish robustness
to general bid adaptation or budget pacing.
\runin{Table 1 structure.} Panel (a) of Table~\ref{tab:main} reports episode-501
clean efficiency $\mathrm{Eff}=\Wel(x)/\Wel(O)$, efficiency after doubling the
strongest non-anchor rival's bid, the improvement over no boost, the episode mean
$\pm$ std over 501--505, and the worst per-policy shock efficiency. Panel (b)
reports per-policy means over 501--505 as CNY and efficiency, with the gap column as
canonical minus the best baseline on that row. Blue and orange mark the best and
second-best entries. Clipped Myerson equals no boost.

\runinfixed{Headline and baselines.} The certified canonical rule attains $98.8\%$
under the twofold shock, versus $92.2\%$ for minimal-flip and $89.0\%$ for no
boost. This shock advantage is the main empirical result. The unclipped Myerson
benchmark reaches $91.9\%$ under shock and is the strongest out-of-class comparator,
but it uses negative offsets, or score demotion, and therefore ranges over a
superset of the nonnegative-boost domain studied here. Among the nonnegative-
boost baselines, minimal-flip is strongest on episode 501 at $92.2\%$. Canonical exceeds
both, by $6.9$ percentage points (pp) over Myerson and $6.6$~pp over minimal-flip on this episode.
BSP-AM~\cite{golrezaei2021boosted} attains $90.5\%$ clean and $87.4\%$ under
shock, below no boost in both columns ($2.1$~pp clean and $1.6$~pp under the
shock). It maximizes the platform's boosted second-price revenue on historical
bids. Its
boost-fitting objective is revenue-only, while the simulator still adds $q$ to
the ranking score.

\subsection{Mechanism diagnostics}
\label{sec:exp_diagnostics}

\runin{Shock and five-episode results.} Under the twofold shock, efficiency
ranges from $94.9$ to $100.0\%$ for canonical and $84.3$ to $99.6\%$ for
minimal-flip, with means of $98.8\%$ and $92.2\%$. Canonical is at least as good
as minimal-flip in all $10/10$ policies, $50/50$ episode-policy cells, and
$4/4$ shock levels $\lambda\in\{1.2,1.5,2,3\}$. The advantage is not specific to
episode 501: over episodes 501--505 the same shock gives canonical
$97.63\pm2.24\%$, minimal-flip $91.12\pm1.81\%$, and no boost $86.19\pm5.50\%$
(mean $\pm$ std across episode means). The paired gap is $6.51\pm0.47$~pp
($t(4)=31.1$, $p<0.001$). Table~\ref{tab:main}(b) reports the per-policy means behind these
aggregates. At $\alpha=0.5$ the cap binds on $96.4\%$ of impressions, so the
certificate acts as a residual allocation buffer under bid drift. Its payment
cost is quantified in Section~\ref{sec:exp_tradeoff}.

\runin{Ranking-signal decomposition.} Table~\ref{tab:bidonly} separates
the two ranking signals. Bid-only ranks by $b$ alone, no boost ranks by $b+q$,
and canonical ranks by $b+q+z$. On episode 501, bid-only
averages $82.1\%$ versus $92.6\%$ for no boost and $99.5\%$ for canonical. Over
episodes 501--505, bid-only averages $81.22\pm3.02\%$ versus $90.47\pm4.50\%$ for
no boost. The quality signal is positive on every policy (e.g., CQL:
$80.6\%$ versus $74.9\%$).

\subsection{Controlled stress test: isolating the certificate effect}
\label{sec:exp_stress}
\begin{figure}[t]
  \centering
  \includegraphics[width=\linewidth]{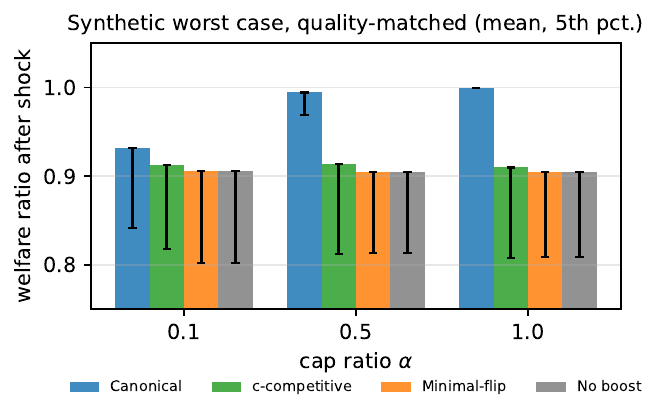}
  \caption{Controlled single-rival stress test. Bars show mean welfare ratios
  after the shock, and lower whiskers mark the fifth percentile.}
  \Description{Bar chart of welfare after a bid shock for the canonical boost, c-competitive, minimal-flip, and no boost at th
  ree cap ratios. Bars show means, with lower whiskers extending to the fifth percentile.}
  \label{fig:stress}
\end{figure}

Table~\ref{tab:main} shows the response to bid shocks on AuctionNet.
We next isolate the source of the gap on random
$\kappa$-bounded instances. The strongest non-anchor rival is raised to the bid cap $\kappa v$ after the
boost is computed on the pre-shock bids, so the test directly probes the
situation for which a rival-independent certificate margin is useful. For a multiplicative shock with
factor $\lambda$, the corresponding additive score increase is
$(\lambda-1)b_{r,j}$. The $\kappa$ projection can only weaken it, so the
reported effect is a lower bound on the unclipped increase. The guarantee is a worst-case floor,
not a prediction of average performance. Figure~\ref{fig:stress} shows the results
over $2{,}000$ draws per cap ratio. At $\alpha=0.5$, the canonical boost
averages $0.995$ with a fifth percentile of $0.969$. The c-competitive
baseline averages $0.914$ with a fifth percentile of $0.812$.
Both minimal-flip and no boost average $0.905$ with a
fifth percentile of $0.814$. We observed no numerical violation of the theorem's
lower bound in these draws. This is a sanity check, not a proof. The
gap between canonical and minimal-flip grows with $\alpha$, reaching $1.000$
versus $0.905$ at $\alpha=1$. The advantage therefore also appears under
controlled conditions, consistent with the certificate margin in
Proposition~\ref{prop:robust}.

The identical outcomes of minimal-flip and no boost in Figure~\ref{fig:stress}
follow from the stress construction. When a boost is needed, minimal-flip
raises the anchor only to the strongest rival's original score, or less if
the boost cap binds. Raising that rival's bid with boosts fixed then leaves
it ahead under both rules. When no boost is needed, the two rules coincide
apart from numerical tie-breaking. Their equality here is therefore specific
to the perturbation, which directly tests whether the boost leaves a
residual score margin.

\subsection{Operating point and payment trade-off}
\label{sec:exp_tradeoff}
\runin{Payment cost and cap sweep.} At $\alpha=0.5$, canonical reaches $99.5\%$ clean welfare and $98.8\%$ under shock, $6.6$~pp above minimal-flip, and retains $132.8$ of the $375.6$ CNY that minimal-flip keeps (Table~\ref{tab:robust}). The cap moves this trade-off: lowering it to $\alpha=0.3$ returns $220.7$ CNY ($+66\%$ retained revenue) while shock efficiency falls by $2.3$ pp to $96.5\%$; lowering it to $\alpha=0.1$ returns another $129$ CNY but gives up $5.0$ pp and nearly all of the advantage over minimal-flip. Raising it to $\alpha=1.0$ gains $1.1$ pp at more than twice the revenue cost per pp of the previous step. We therefore call $\alpha=0.5$ a practical operating point, not a derived optimum, and leave the choice of point on this curve to the platform.

\subsection{Stability, sensitivity, and scope}
\label{sec:exp_scope}
\runin{Stability and sensitivity.} Appendix~\ref{app:extra} reports the
five-episode check, revenue, implementation details, and the full per-policy
matrix. Canonical and minimal-flip remain at the top in the $\xi$ and shock
sweeps, with the canonical advantage peaking at $7.1$~pp at $\lambda=1.5$.
The rival-independent margin provides protection against bid drift at
$35$--$53\%$ of minimal-flip's retained revenue on episode 501 (clean versus twofold shock). A platform that trusts its bid
stream can use minimal-flip.

\section{Conclusion}
\label{sec:conclusion}
We study boost design with quality externalities under a rule that deducts
boosts from payments. The capped certificate yields a welfare lower bound
and, when its margin seats the welfare-optimal winner, a closed-form
revenue-maximal rule. On AuctionNet, canonical retains $98.8\%$ under a
twofold rival shock versus $92.2\%$ for minimal-flip ($97.6\%$ versus $91.1\%$
over five episodes, or $90.2\%$ for a Myerson comparator with negative boosts),
at $35$--$53\%$ of minimal-flip's retained revenue on episode 501. The evidence is offline,
single-slot, fixed-bid, with a frozen quality stream and a calibrated quality
scale. Multi-slot allocation, strategic adaptation, and learned bids are future
work.

\clearpage
\bibliographystyle{ACM-Reference-Format}
\bibliography{ref}

\appendix

\section{Proof of Theorem~\ref{thm:main}}
\label{app:thm1}

\noindent We prove the two inequalities of the theorem. Throughout,
$A=\{j:w(j)=o(j)\}$.

For $j\notin A$ the benchmark winner $o(j)$ is a losing bidder, so
$s_{(2),j}\ge s_{o(j),j}$. Combined with the margin $z_{o,j}-z_{w,j}\ge m_j$ and
with $[a]^+\ge a$, this gives
\begin{equation}\label{eq:pf1}
\begin{aligned}
  p_{w,j}\ &\ge\ s_{(2),j}-q_{w,j}-z_{w,j}
           \ \ge\ b_{o,j}+q_{o,j}+z_{o,j}-q_{w,j}-z_{w,j}\\
           &\ge\ b_{o,j}+q_{o,j}+m_j-q_{w,j}.
\end{aligned}
\end{equation}
In \eqref{eq:pf1} the first inequality is $[a]^+\ge a$. The second is
$s_{(2),j}\ge s_{o(j),j}=b_{o,j}+q_{o,j}+z_{o,j}$, which holds because $o(j)$
loses when $j\notin A$. The third is the margin $z_{o,j}-z_{w,j}\ge m_j$.
Summing \eqref{eq:pf1} over $j\notin A$ and dividing by $c+1$,
\begin{equation}\label{eq:pf2}
  \frac{\Rev(x)}{c+1}\ \ge\ \frac{1}{c+1}\!\!\sum_{j\notin A}\!(b_{o,j}+q_{o,j}+m_j)
  \ -\ \frac{1}{c+1}\!\!\sum_{j\notin A}\!q_{w,j}.
\end{equation}

Adding $\Wel(x)=\sum_i\min\{B_i,V_i(x)\}+\sum_{j\in A}q_{o,j}+\sum_{j\notin A}q_{w,j}$
to \eqref{eq:pf2}, using
$(v_{o,j}+q_{o,j})\le \frac{b_{o,j}+q_{o,j}+m_j}{c+1}+\delta_j$, and collecting
$\sum_{j\in A}q_{o,j}+\sum_{j\notin A}q_{o,j}=\sum_j q_{o,j}$, yields
\begin{align}\label{eq:pf3}
  \Wel(x)+\frac{\Rev(x)}{c+1}
  \ \ge\ & \Big[\sum_i\min\{B_i,V_i(x)\}+\sum_{j\notin A}v_{o,j}\Big] \nonumber \\
  &+\sum_j q_{o,j}+\frac{c}{c+1}\!\!\sum_{j\notin A}\!q_{w,j}
  -\sum_{j\notin A}\delta_j .
\end{align}
The bracketed term in \eqref{eq:pf3} is the value/budget contribution, whereas
the remaining quality terms split $\sum_j q_{o,j}$ into a directly counted part
$\sum_{j\in A}q_{o,j}$ and the residual $\frac{c}{c+1}\sum_{j\notin A}q_{w,j}$.
To estimate the bracketed term, set for each advertiser $i$
\[
\begin{aligned}
  G_i&=\!\!\sum_{j\in A}\!v_{i,j}\mathbf 1\{o(j)=i\},\qquad
  P_i=\!\!\sum_{j\notin A}\!v_{i,j}\mathbf 1\{o(j)=i\},\\
  X_i&=\!\!\sum_{j\notin A}\!v_{i,j}\mathbf 1\{w(j)=i\}.
\end{aligned}
\]
Since $j\in A$ means $w(j)=o(j)$, $V_i(O)=G_i+P_i$ and $V_i(x)=G_i+X_i$, so
\begin{equation}\label{eq:pf4}
\begin{aligned}
  \min\{B_i,V_i(O)\}&-\min\{B_i,V_i(x)\}\\
  &=\ \min\{B_i,G_i+P_i\}-\min\{B_i,G_i+X_i\}\\
  &\le\ (P_i-X_i)^+\ \le\ P_i .
\end{aligned}
\end{equation}
In \eqref{eq:pf4} the equality substitutes $V_i(O)=G_i+P_i$ and
$V_i(x)=G_i+X_i$. The second inequality is the non-decreasing, $1$-Lipschitz
property of $u\mapsto\min\{B_i,u\}$. The last is $X_i\ge0$. Summing
\eqref{eq:pf4} over $i$ and using $\sum_i P_i=\sum_{j\notin A}v_{o,j}$,
\begin{equation}\label{eq:pf5}
  \sum_i\min\{B_i,V_i(x)\}+\sum_{j\notin A}v_{o,j}\ \ge\ \sum_i\min\{B_i,V_i(O)\}
  \ =\ \Wel(O)-\sum_j q_{o,j}.
\end{equation}
Substituting \eqref{eq:pf5} into \eqref{eq:pf3} and dropping the non-negative
residual $\frac{c}{c+1}\sum_{j\notin A}q_{w,j}$,
\begin{equation}\label{eq:pf6}
  \Wel(x)+\frac{\Rev(x)}{c+1}\ \ge\ \Wel(O)-\sum_{j\notin A}\delta_j .
\end{equation}
Under Assumption~\ref{ass:kappa}, $b_{i,j}\le\kappa v_{i,j}$ together with the
budget constraint in \eqref{eq:adv} gives, per advertiser,
$\sum_j p_{i,j}\le\min\{B_i,\kappa V_i(x)\}\le\kappa\min\{B_i,V_i(x)\}$ (as
$\kappa\ge1$), hence $\Rev(x)\le\kappa\,\Wel_c(x)\le\kappa\,\Wel(x)$. So
$\Wel(x)+\frac{\Rev(x)}{c+1}\le\frac{c+1+\kappa}{c+1}\Wel(x)$. With \eqref{eq:pf6},
\begin{equation}\label{eq:pf7}
  \Wel(x)\ \ge\ \frac{c+1}{c+1+\kappa}\Big(\Wel(O)-\sum_{j\notin A}\delta_j\Big),
\end{equation}
the first inequality of the theorem.

It remains to bound $\sum_{j\notin A}\delta_j$. If $M_j\le Z_{\max}$ then
$m_j=M_j$ and
$b_{o,j}+q_{o,j}+M_j=(c+1)(v_{o,j}+q_{o,j})+[\max\{b_{o,j},v_{o,j}\}-v_{o,j}]
\ge(c+1)(v_{o,j}+q_{o,j})$ by $b_{o,j}+(v_{o,j}-b_{o,j})^+=\max\{b_{o,j},v_{o,j}\}$,
so $\delta_j=0$. If $M_j>Z_{\max}$ then $m_j=Z_{\max}$ and
$\delta_j=\frac{1}{c+1}\POW{(c+1)(v_{o,j}+q_{o,j})-b_{o,j}-q_{o,j}-Z_{\max}}$,
hence, since only impressions with $M_j>Z_{\max}$ have $\delta_j\neq0$,
\begin{align}\label{eq:pf8}
  \sum_{j\notin A}\delta_j\ &\le\ \sum_{j}\delta_j \nonumber \\
  &=\frac{1}{c+1}\!\!\!\sum_{\substack{j:\,M_j>Z_{\max}}}\!\!\!  
   \POW{(c+1)(v_{o,j}+q_{o,j})-b_{o,j}-q_{o,j}-Z_{\max}} \nonumber \\
  &=\frac{c+2}{c+1}\,\Lambda(Z_{\max}).
\end{align}
Substituting \eqref{eq:pf8} into \eqref{eq:pf7} gives the second inequality of
the theorem. Increasing $Z_{\max}$ shrinks $\{j:M_j>Z_{\max}\}$ and decreases each
summand, so $\Lambda$ is non-increasing, $\rho_\kappa$ is non-decreasing, and
$\rho_\kappa=\frac{c+1}{c+1+\kappa}$ once $Z_{\max}\ge\max_j M_j$. \qed

\medskip
\runin{Proof of Corollary~\ref{cor:rev}.} The first inequality of
\eqref{eq:correv} is \eqref{eq:pf6} with $\Rev(x)/(c+1)$ moved to the right-hand
side. The second follows because $\delta_j\ge0$ for every $j$ and $\delta_j=0$
whenever $M_j\le Z_{\max}$, so that $\sum_{j\notin A}\delta_j\le\sum_j\delta_j
=D(Z_{\max})$. The identity $D(Z_{\max})=\frac{c+2}{c+1}\Lambda(Z_{\max})$ is
immediate from the definitions of $\delta_j$ and $\Lambda$. \qed

\medskip
\runin{Proof of Lemma~\ref{lem:couple}.} Since $w(j)$ is the winner,
$s_{(2),j}\le s_{w,j}=b_{w,j}+q_{w,j}+z_{w,j}$. Therefore
\begin{align}
p_{w,j}=\POW{s_{(2),j}-q_{w,j}-z_{w,j}}
  \ &\le\ \POW{b_{w,j}+q_{w,j}+z_{w,j}-q_{w,j}-z_{w,j}} \nonumber \\
  \ &=\ \POW{b_{w,j}}\ =\ b_{w,j}, 
\end{align}

using $b_{w,j}\ge0$ and the monotonicity of $[\cdot]^+$. Moreover, whenever the
clamp is inactive ($s_{(2),j}>q_{w,j}+z_{w,j}$) we have
$p_{w,j}=s_{(2),j}-q_{w,j}-z_{w,j}$, which is affine with
$\partial p_{w,j}/\partial q_{w,j}=\partial p_{w,j}/\partial z_{w,j}=-1$: each
extra unit of quality or boost is exactly one unit of forgone payment. \qed

\medskip
\runin{Proof of Corollary~\ref{cor:imposs}.} Fix
$\delta\in(0,\tfrac12)$, $\kappa\ge1$, and $\varepsilon>0$. Take a single
impression with two advertisers: advertiser~$1$ (value-rich) has
$v_1=V:=(1-2\delta)(w+Q)$, $q_1=0$, $B_1=\infty$. Advertiser~$2$ (quality-rich) has
$v_2=w$, $q_2=Q$, and $B_2=w$, where $Q>0$ and $w>0$ is small enough that
$\kappa w<\varepsilon(w+Q)$ (possible for every finite $\kappa$). With one
impression an outcome admits either advertiser:
seating~$1$ gives $\Wel=\min\{B_1,v_1\}+q_1=V=(1-2\delta)(w+Q)$, while seating~$2$
gives $\Wel=\min\{B_2,v_2\}+q_2=w+Q$. Hence
$\Wel(O)=\max\{V,w+Q\}=w+Q$ (as $1-2\delta<1$), and whenever
$\Wel(x)\ge(1-\delta)\Wel(O)$, advertiser~$2$ is seated, because seating~$1$
yields only $(1-2\delta)\Wel(O)<(1-\delta)\Wel(O)$. When advertiser~$2$ wins, the bid cap and payment rule give
$p_2\le b_2\le\kappa v_2=\kappa w$. Hence
$\Rev(x)=p_2\le\kappa w<\varepsilon(w+Q)=\varepsilon\,\Wel(O)$. Since the argument
holds for an arbitrary boost $z$, the claim follows. \qed

\medskip
\runin{Proof of Corollary~\ref{cor:ref}.} The proof of
Theorem~\ref{thm:main} uses the reference allocation $O=\argmax_x\Wel(x)$ only
through three properties, none of which uses optimality:
(i) for $j\notin A$, the reference winner $o(j)$ loses, so
$s_{(2),j}\ge s_{o(j),j}$, and the margin
$z_{o(j),j}-z_{i,j}\ge m_j$ of Definition~\ref{def:bc} holds.
(ii) the per-advertiser decomposition of $V_i$ into the part on
$A=\{j:w(j)=o(j)\}$ and its complement, used in \eqref{eq:pf4}.
(iii) the benchmarking identity
$\sum_i\min\{B_i,V_i(O)\}=\Wel(O)-\sum_j q_{o(j),j}$ in \eqref{eq:pf5}.
Replace $O$ by an arbitrary allocation $R$ with per-impression winner $o_R(j)$ and
impose Definition~\ref{def:bc} with $o_R$ in place of $o$ (so that property (i)
holds with $m_j$ as defined in the statement). Then properties (ii)--(iii) hold
verbatim with $o\to o_R$, $A\to A_R:=\{j:w(j)=o_R(j)\}$, and
$\Wel(O)\to\Wel(R)$. Running the chain \eqref{eq:pf1}--\eqref{eq:pf6} unchanged
therefore yields
$\Wel(x)+\frac{\Rev(x)}{c+1}\ge\Wel(R)-\sum_{j\notin A_R}\delta_j^{R}$. Combining
with $\Rev(x)\le\kappa\,\Wel(x)$ (same argument as in the proof of
Theorem~\ref{thm:main}) gives
$\Wel(x)\ge\frac{c+1}{c+1+\kappa}(\Wel(R)-\sum_{j\notin A_R}\delta_j^{R})
\ge\frac{c+1}{c+1+\kappa}\Wel(R)-\frac{c+2}{c+1+\kappa}\Lambda_R(Z_{\max})$. \qed
\section{Additional experiments}
\label{app:extra}

\begin{sloppypar}
\runin{Environment.} Experiments are CPU-only (Python~3.12.1, NumPy~2.3.3,
PyTorch~2.13.0+cpu) and deterministic given the episode index: episode $e$
seeds the environment with $42+e$, and at tick $t$ with $N_t$ impressions the
quality draw is $\mathrm{RandomState}(42+1000e+10t+N_t)$, clipped to $[0,50]$
units and priced at $3.56\%$ of the mean commercial value. One episode-policy
takes 8--15\,s, so a ten-policy sweep costs 2--3 minutes per mechanism.



\runin{Shock and metrics.} The shock is applied after $z$: with
$r_j=\argmax_{i\ne o(j)}(b_{i,j}+\xi q_{i,j})$, set
$b_{r_j,j}\leftarrow\lambda b_{r_j,j}$ and re-project
$b\leftarrow\min\{b,\kappa v\}$, with $\lambda=2$ unless swept. The auction is
single-slot second score with $s=b+z+\xi q$. The winner pays
$p_w=[s_{(2)}-q_w-z_w]^+$ at zero reserve, not a BSP division. Let $x_t$ be the
allocation at tick $t$ and
$\mathrm{Wel}_t=\sum_i\min\{B_i,V_i(x_t)\}+\xi\sum_j q_{w(j),j}$. Let $O_t$ be the
reference allocation at tick $t$, whose winner on impression $j$ is $o(j)$, so
that $\mathrm{Wel}(O_t)$ is the same liquid-welfare functional evaluated at $O_t$.
Efficiency is $\sum_t\mathrm{Wel}_t/\sum_t\mathrm{Wel}(O_t)$, the ratio of sums reported in all tables, and
revenue is $\sum_t\sum_j p_{w(j),j}$.

\runin{Parameters and baselines.} We fix $c=2$, $\xi=1$, $\kappa=2.5$, and
$\alpha=0.5$. The pre-trained bidders exceed $v$ on $67.5\%$ of
impression-bidder pairs. The c-competitive baseline uses $c=1.5$ as in
\cite{deng2021towards}. Uniform adds an untuned $0.3$. BSP-AM fits multipliers
$\beta_i$ by coordinate descent on the previous tick's bids (proximal penalty
$0.01$, at most 200 sweeps, stopping when
$\sum_i(\beta_i^{(k)}-\beta_i^{(k-1)})^2\le0.01$, normalized so
$\min_i\beta_i=1$), using the last 50 auctions. The fitted factors give
$z_{i,j}=(\beta_i-1)b_{i,j}$ with $q$ omitted from the fit. Myerson computes per
impression the empirical virtual value of $v_{i,j}+q_{i,j}$ from three
equal-width histogram bins with empirical distribution $F$ and density $f$, uses
$\phi(x)=x-(1-F(x))/f(x)$, and sets
$z_{i,j}=\phi(v_{i,j}+q_{i,j})-(v_{i,j}+q_{i,j})$. The unclipped variant keeps
negative offsets and the clipped one truncates them at zero. Both are
heuristic, non-certified baselines. All mechanisms share one evaluation
interface.

\runin{Verification.} As an arithmetic anchor, \eqref{eq:canonical} on the
episode-501 state with CQL yields $\mathrm{Wel}=458.37$ and
$\mathrm{Rev}=70.51$, i.e.\ $99.98\%$ of $\Wel(O)=458.48$.
\end{sloppypar}

\runin{Multi-episode stability.} Table~\ref{tab:episodes} reports mean
efficiency over episodes 501--505. Canonical and minimal-flip remain ahead of
uniform and no boost. Canonical's advantage appears under the bid shock.
Over the five episodes, the canonical boost averages $99.40\pm0.23\%$,
minimal-flip $99.42\pm0.24\%$, uniform $95.27\pm0.65\%$, and no boost
$90.47\pm4.50\%$. Minimal-flip has the highest clean-state point estimate.
The paired gap, canonical minus minimal-flip, is
$-0.02\pm0.01$ pp (paired $t(4)\approx-2.41$, $p\approx0.07$).
The $0.02$~pp difference is practically negligible and not significant at the
$5\%$ level.

\begin{table}[t]
\centering
\footnotesize
\caption{Mean efficiency over episodes 501--505, averaged over the ten
bidding policies. $\pm$ is one standard deviation across the five episode
means (each episode mean is over the ten policies), not across the 50
episode-policy cells. The per-policy range is reported in
Table~\ref{tab:fullmatrix}. Clipped Myerson coincides with no boost, and the
unclipped penalty variant is the Myerson row. Best entry is
shaded blue in bold, the second best orange with an underline, and ties share
the mark.}
\label{tab:episodes}
\begin{tabular}{@{}l ccc@{}}
\toprule
\rowcolor{headgray}
Mechanism & Mean $\pm$ std & Min & Max \\
\midrule
Canonical boost & \usecond{99.40\% $\pm$ 0.23\%} & \bfirst{99.00\%}{} & \bfirst{99.56\%}{} \\
c-competitive & 92.29\% $\pm$ 1.55\% & 90.96\% & 94.85\% \\
Minimal-flip & \bfirst{99.42\% $\pm$ 0.24\%}{} & \bfirst{99.00\%}{} & \bfirst{99.56\%}{} \\
Uniform & 95.27\% $\pm$ 0.65\% & \usecond{94.76\%} & \usecond{96.41\%} \\
Myerson (unclipped) & 93.48\% $\pm$ 2.40\% & 89.20\% & 94.83\% \\
No boost & 90.47\% $\pm$ 4.50\% & 82.47\% & 93.01\% \\
Clipped Myerson & 90.47\% $\pm$ 4.50\% & 82.47\% & 93.01\% \\
BSP-AM & 88.13\% $\pm$ 4.66\% & 79.89\% & 90.81\% \\
\midrule
Gap vs. strongest (pp) & \cost{$-0.02 \pm 0.01$} & \cost{$-0.03$} & \cost{$0.00$} \\
\bottomrule
\end{tabular}
\end{table}

\begin{table}[t]
\centering
\footnotesize
\caption{Operating point and payment trade-off on episode 501, averaged over
the ten bidding policies. $\alpha$ scales the cap,
$Z_{\max}=\alpha\max_{i,j}(v+q)$. ``Shock'' raises the strongest non-anchor
rival's bid by $2\times$, applied after the boost is computed on the pre-shock
bids and after the ROI projection, then re-clipped to $\kappa v$ before the
auction. ``Rev. C/M'' is clean retained revenue in CNY per
episode for canonical/minimal-flip. The gap is canonical minus
minimal-flip. The guarantee $\rho_\kappa$ is computed
on the pre-shock state, instance by instance, and averaged over the ten policies. The theoretical
maximum is $(c+1)/(c+1+\kappa)=3/5.5=0.545$.}
\label{tab:robust}
\setlength{\tabcolsep}{3.5pt}
\begin{tabular}{@{}l cccccc@{}}
\toprule
\rowcolor{headgray}
$\alpha$ & Bind & $\rho_\kappa$ & Clean Eff & Shock Eff & Gap vs. min.-flip (pp) & Rev. C/M \\
\midrule
0.1 & 1.00 & 0.264 & 97.1\% & 91.5\% & +1.1 & 350/402 \\
0.3 & 0.97 & 0.317 & 99.0\% & 96.5\% & +4.8 & 221/386 \\
0.5 & 0.96 & 0.367 & 99.5\% & 98.8\% & +6.6 & 133/376 \\
1.0 & 0.89 & 0.473 & 100.0\% & 99.9\% & +7.4 & 29/366 \\
2.0 & 0.04 & 0.544 & 100.0\% & 100.0\% & +7.4 & 1/364 \\
\bottomrule
\end{tabular}
\end{table}

\begin{table}[t]
\centering
\footnotesize
\caption{Per-policy clean welfare efficiency on episode 501
($\kappa=2.5$, $c=2$, $\alpha=0.5$), relative to the common benchmark
$\Wel(O)=458.48$ CNY. Shock results are in Table~\ref{tab:main}. Shading is as in
Table~\ref{tab:episodes}. The gap is canonical minus minimal-flip.}
\label{tab:fullmatrix}
\scriptsize
\setlength{\tabcolsep}{1pt}
\begin{tabular}{@{}l ccccccc r@{}}
\toprule
\rowcolor{headgray}
\textbf{Policy} & \textbf{Canonical} & \textbf{c-comp.} & \textbf{Minimal} & \textbf{Myerson} & \textbf{BSP-AM} & \textbf{Uniform} & \textbf{No boost} & \shortstack{\textbf{Gap vs.}\\\textbf{min.-flip (pp)}} \\
\midrule
PID & \bfirst{100.0\%}{} & 91.4\% & \bfirst{100.0\%}{} & 95.7\% & 91.8\% & \usecond{97.3\%} & 94.9\% & 0.0 \\
OnlineLP & \bfirst{100.0\%}{} & 91.4\% & \bfirst{100.0\%}{} & \bfirst{100.0\%}{} & 95.3\% & \usecond{99.7\%} & \bfirst{100.0\%}{} & 0.0 \\
BC & \bfirst{100.0\%}{} & 91.4\% & \bfirst{100.0\%}{} & \usecond{97.6\%} & 92.8\% & 97.4\% & 97.5\% & 0.0 \\
Combo & \bfirst{100.0\%}{} & 91.4\% & \bfirst{100.0\%}{} & 93.7\% & 91.0\% & \usecond{97.0\%} & 92.0\% & 0.0 \\
BCQ & \bfirst{100.0\%}{} & 91.4\% & \bfirst{100.0\%}{} & 93.7\% & 92.2\% & \usecond{97.2\%} & 92.1\% & 0.0 \\
CQL & \bfirst{100.0\%}{} & \usecond{92.0\%} & \bfirst{100.0\%}{} & 91.4\% & 83.5\% & 86.7\% & 80.6\% & 0.0 \\
TD3BC & \bfirst{100.0\%}{} & 91.4\% & \bfirst{100.0\%}{} & \usecond{96.9\%} & 92.3\% & 96.8\% & 96.7\% & 0.0 \\
IQL & \bfirst{100.0\%}{} & 91.6\% & \bfirst{100.0\%}{} & 93.7\% & 90.0\% & \usecond{93.9\%} & 92.9\% & 0.0 \\
Mopo & \bfirst{100.0\%}{} & 91.4\% & \bfirst{100.0\%}{} & 93.7\% & 92.1\% & \usecond{97.2\%} & 92.3\% & 0.0 \\
Mix & \usecond{95.4\%} & 90.4\% & \bfirst{95.7\%}{} & 88.9\% & 84.4\% & 86.9\% & 86.9\% & \cost{$-0.3$} \\
\midrule
\textbf{Mean} & \usecond{99.5\%} & 91.4\% & \bfirst{99.6\%}{} & 94.5\% & 90.5\% & 95.0\% & 92.6\% & \cost{$-0.1$} \\
\bottomrule
\end{tabular}
\end{table}

\begin{table}[t]
\centering
\footnotesize
\scriptsize
\caption{Shock-magnitude sweep on episode 501 (ten-policy mean, $\alpha=0.5$).
The strongest non-anchor rival's bid is multiplied by $\lambda$ after the boost
is computed on the pre-shock bids. The shocked bid is re-clipped to
$\kappa v$. Shading is as in Table~\ref{tab:episodes}. The gap is canonical minus
minimal-flip.}
\label{tab:lambda}
\setlength{\tabcolsep}{2pt}
\begin{tabular}{@{}l cccccc c r@{}}
\toprule
\rowcolor{headgray}
$\lambda$ & Canonical & c-comp. & Minimal-flip & Myerson & BSP-AM & Uniform & No boost & \shortstack{Gap vs.\\min.-flip (pp)} \\
\midrule
1.2 & \bfirst{99.5\%}{} & 91.4\% & \usecond{93.4\%} & 92.9\% & 89.1\% & 92.1\% & 90.2\% & \gain{+6.1} \\
1.5 & \bfirst{99.4\%}{} & 91.4\% & \usecond{92.3\%} & 92.1\% & 88.3\% & 89.9\% & 89.1\% & \gain{+7.1} \\
2.0 & \bfirst{98.8\%}{} & 91.4\% & \usecond{92.2\%} & 91.9\% & 87.4\% & 89.4\% & 89.0\% & \gain{+6.6} \\
3.0 & \bfirst{98.1\%}{} & 91.5\% & \usecond{92.2\%} & 91.8\% & 86.6\% & 89.4\% & 89.0\% & \gain{+5.9} \\
\bottomrule
\end{tabular}
\end{table}

\begin{table}[t]
\centering
\footnotesize
\caption{Ranking-signal decomposition on episode 501 (frozen quality stream,
$\kappa=2.5$, $c=2$, $\alpha=0.5$). Bid-only ranks by $b$ alone. No boost
ranks by $b+q$, and canonical ranks by $b+q+z$. The bid-only vs.\ no-boost gap
isolates the quality signal. The no-boost vs.\ canonical gap isolates the boost.
Shading is as in Table~\ref{tab:episodes}. The gap is canonical minus the
stronger of the two signal-only rules.}
\label{tab:bidonly}
\begin{tabular}{@{}l ccc r@{}}
\toprule
\rowcolor{headgray}
Policy & Bid-only & No boost & Canonical & Gap vs. signal-only (pp) \\
\midrule
PID & 85.5\% & \usecond{94.9\%} & \bfirst{100.0\%}{} & \gain{+5.1} \\
OnlineLP & \usecond{91.2\%} & \bfirst{100.0\%}{} & \bfirst{100.0\%}{} & 0.0 \\
BC & 90.6\% & \usecond{97.5\%} & \bfirst{100.0\%}{} & \gain{+2.5} \\
Combo & 69.0\% & \usecond{92.0\%} & \bfirst{100.0\%}{} & \gain{+8.0} \\
BCQ & 76.9\% & \usecond{92.1\%} & \bfirst{100.0\%}{} & \gain{+7.9} \\
CQL & 74.9\% & \usecond{80.6\%} & \bfirst{100.0\%}{} & \gain{+19.4} \\
TD3BC & 89.4\% & \usecond{96.7\%} & \bfirst{100.0\%}{} & \gain{+3.3} \\
IQL & 84.7\% & \usecond{92.9\%} & \bfirst{100.0\%}{} & \gain{+7.1} \\
Mopo & 76.6\% & \usecond{92.3\%} & \bfirst{100.0\%}{} & \gain{+7.7} \\
Mix & 82.1\% & \usecond{86.9\%} & \bfirst{95.4\%}{} & \gain{+8.5} \\
\midrule
\textbf{Mean} & 82.1\% & \usecond{92.6\%} & \bfirst{99.5\%}{} & \gain{+6.9} \\
\bottomrule
\end{tabular}
\end{table}

\begin{table}[t]
\centering
\footnotesize
\scriptsize
\caption{Quality-weight sensitivity on the three-policy subset CQL, IQL, and
TD3BC (episode 501, $\kappa=2.5$, $c=2$, $\alpha=0.5$). Absolute
efficiency changes because the benchmark weights quality differently. Canonical
and minimal-flip remain at the top. Shading is as in Table~\ref{tab:episodes}.
The gap is canonical minus minimal-flip.}
\label{tab:xi}
\setlength{\tabcolsep}{2pt}
\begin{tabular}{@{}l cccccc c r@{}}
\toprule
\rowcolor{headgray}
$\xi$ & Canonical & c-comp. & Minimal-flip & Myerson & BSP-AM & Uniform & No boost & \shortstack{Gap vs.\\min.-flip (pp)} \\
\midrule
0.1 & \usecond{99.83\%} & 99.67\% & \bfirst{99.86\%}{} & 94.20\% & 82.54\% & 94.52\% & 84.09\% & \cost{$-0.03$} \\
1.0 & \bfirst{99.99\%}{} & 91.65\% & \bfirst{99.99\%}{} & \usecond{94.02\%} & 88.58\% & 92.48\% & 90.11\% & 0.00 \\
10.0 & \bfirst{100.00\%}{} & 96.86\% & \bfirst{100.00\%}{} & 98.91\% & 99.11\% & 99.30\% & \usecond{99.46\%} & 0.00 \\
\bottomrule
\end{tabular}
\end{table}

\runin{Revenue.} At $\alpha=0.5$, episode-501 mean retained revenue is $132.8$
CNY for canonical, $375.6$ for minimal-flip, $396.7$ for uniform, and $394.6$ for
no boost. Over episodes 501--505, canonical and minimal-flip retain $174.4$ and
$452.4$ CNY. Canonical retains $35\%$ of minimal-flip's revenue on episode 501
($53\%$ under the twofold shock) and $39\%$ over the five episodes ($62\%$
under the shock).

\runin{Implementation details.} Clipped Myerson equals no boost in all 50
episode-policy cells. Unclipped Myerson keeps negative offsets
(Table~\ref{tab:main}). Canonical and minimal-flip use \eqref{eq:canonical} and
\eqref{eq:flip}, with $10^{-9}$ used only to break ties.

\runin{Certificate strength.} Table~\ref{tab:c} varies $c\in\{0,1,2,4\}$ at the
operating cap and at a nearly slack cap. At $\alpha=0.5$ the cap already binds
on $89$--$98\%$ of impressions for $c\ge1$, so $m_j=Z_{\max}$ and the realized
efficiency is flat ($99.53\%$), while $\rho_\kappa$ first rises and
then falls ($0.286\to0.397\to0.367\to0.277$). A larger $c$ inflates the
required margin $M_j$ and hence $\Lambda$. At $\alpha=2$ both
$\rho_\kappa$ and the realized efficiency increase with $c$, but retained
revenue falls monotonically from $392.0$ at $c=0$ to $1.2$ CNY at $c=4$. Thus
$c=2$ is not tuned for realized performance but follows from a target guarantee.
When $\kappa=1$ and the cap does not bind, the theorem takes the constant form
$\Wel(x)\ge\frac{c+1}{c+2}\Wel(O)$, so a platform that targets $75\%$ of the
benchmark solves $\frac{c+1}{c+2}=0.75$ and obtains $c=2$. At $\kappa=2.5$ the
operative constant is $3/5.5\simeq0.545$.

\begin{table}[t]
\centering
\footnotesize
\setlength{\tabcolsep}{3.5pt}
\caption{\textbf{Certificate strength.} Episode 501, ten-policy mean. $c$ is
the margin coefficient. $\rho_\kappa$ is the theorem's
floor averaged over the ten policies, Eff.\ the
realized welfare efficiency, and Rev.\ the retained
revenue (CNY). $\alpha=0.5$ is the operating cap and $\alpha=2$ a nearly slack
one.}
\label{tab:c}
\begin{tabular}{@{}l cc cc cc@{}}
\toprule
\rowcolor{headgray}
& \multicolumn{3}{c}{$\alpha=0.5$ (operating)} & \multicolumn{3}{c}{$\alpha=2$ (slack)}\\
\cmidrule(lr){2-4}\cmidrule(lr){5-7}
$c$ & $\rho_\kappa$ & Eff. & Rev. & $\rho_\kappa$ & Eff. & Rev. \\
\midrule
0 & 0.286 & 95.34\% & 392.0 & 0.286 & 95.34\% & 392.0 \\
1 & \textbf{0.397} & 99.53\% & 136.4 & 0.444 & 99.97\% & 64.0 \\
2 & 0.367 & 99.53\% & 132.8 & 0.544 & 100.00\% & 1.2 \\
4 & 0.277 & 99.53\% & 132.8 & \textbf{0.548} & 100.00\% & 1.2 \\
\bottomrule
\end{tabular}
\end{table}

\begin{table}[t]
\centering
\footnotesize
\setlength{\tabcolsep}{3.5pt}
\caption{\textbf{Bid-cap sensitivity under the $2\times$ shock.} Episode 501,
ten-policy mean welfare efficiency and retained revenue (CNY, for
canonical/minimal-flip). $\rho_\kappa$ is the theorem's floor computed on the
pre-shock state and averaged over the ten policies. The clean-state aggregate differs by at most $0.01$
because the shock perturbs the spending path. Shading as in
Table~\ref{tab:episodes}. The gap is canonical minus minimal-flip.}
\label{tab:kappa}
\scriptsize
\setlength{\tabcolsep}{3pt}
\begin{tabular}{@{}l cc cccc c r@{}}
\toprule
\rowcolor{headgray}
$\kappa$ & $\rho_\kappa$ & Canonical & Minimal-flip & Myerson & BSP-AM & No boost & Rev.\ C/M & \shortstack{Gap vs.\\min.-flip (pp)} \\
\midrule
1.0 & 0.311 & \bfirst{100.00\%}{} & \usecond{98.72\%} & 98.13\% & 94.22\% & 98.03\% & 11/215 & \gain{+1.28} \\
1.5 & 0.333 & \bfirst{99.98\%}{} & \usecond{95.66\%} & 94.63\% & 91.00\% & 93.95\% & 52/287 & \gain{+4.32} \\
2.0 & 0.352 & \bfirst{99.51\%}{} & \usecond{93.57\%} & 93.06\% & 88.89\% & 91.23\% & 132/344 & \gain{+5.94} \\
2.5 & 0.367 & \bfirst{98.83\%}{} & \usecond{92.22\%} & 91.85\% & 87.43\% & 89.03\% & 214/401 & \gain{+6.61} \\
3.0 & 0.378 & \bfirst{97.88\%}{} & \usecond{90.93\%} & 90.64\% & 86.40\% & 87.01\% & 295/451 & \gain{+6.95} \\
\bottomrule
\end{tabular}
\end{table}

\runin{Bid-cap sensitivity.} Table~\ref{tab:kappa} sweeps $\kappa$ under the
$2\times$ shock. A looser cap leaves more room for rival overbidding: the
advantage grows from $1.3$~pp at $\kappa=1$ to $7.0$~pp at $\kappa=3$ while the
guarantee rises from $0.311$ to $0.378$, as the cap penalty shrinks faster than
$(c+1)/(c+1+\kappa)$ weakens. Canonical retains $1.5$--$19\times$ less revenue
for this extra margin. The two rules differ by at most $0.04$~pp in the clean
state, and no boost falls from $98.0\%$ to $87.0\%$ under the shock.

\end{document}